\documentclass{ar-1col}

\usepackage[comma]{natbib}
\usepackage{url}
\usepackage{color} 
\usepackage{amsmath} 
\usepackage{amssymb} 

\jname{Xxxx. Xxx. Xxx. Xxx.}
\jvol{AA}
\jyear{YYYY}
\doi{10.1146/((please add article doi))}

\newcommand{\xiion}{$\xi_\mathrm{ion}$}
\newcommand{\fesc}{$f_\mathrm{esc}$}
\newcommand{\heii}{He\,{II}}

\begin{document}

\markboth{Eldridge \& Stanway}{Massive stars and early galaxies}

\title{New Insights into the Evolution of Massive Stars and Their Effects on Our Understanding of Early Galaxies}

\author{Jan J. Eldridge,$^1$ and Elizabeth R. Stanway,$^2$
\affil{$^1$ Department of Physics, University of Auckland, Auckland, New Zealand; email: j.eldridge@auckland.ac.nz}
\affil{$^2$ Department of Physics, University of Warwick, Gibbet Hill Road, Coventry, United Kingdom, CV4 7AL; email: e.r.stanway@warwick.ac.uk}
}

\begin{abstract}

The observable characteristics and subsequent evolution of young stellar populations is dominated by their massive stars. As our understanding of those massive stars and the factors affecting their evolution improves, so our interpretation of distant, unresolved stellar systems can also advance. As observations increasingly probe the distant Universe, and the rare low metallicity starbursts nearby, so the opportunity arises for these two fields to complement one another, and lead to an improved conception of both stars and galaxies. Here we review the current state of the art in modelling of massive star dominated stellar populations, and discuss their applications and implications for interpreting the distant Universe. Our principle findings include:

\smallskip

\begin{minipage}{0.78\textwidth}
\begin{itemize}

\item Binary evolutionary pathways must be included to understand the stellar populations in early galaxies.

\item Observations constraining the extreme ultraviolet spectrum of early galaxies are showing that current models are incomplete. The best current guess is that some form of accretion onto compact remnants is required.

\item The evolution and fates of very massive stars, of the order of 100~M$_{\odot}$ and above, may be key to fully understand aspects of early galaxies.
\end{itemize}
\end{minipage}
\end{abstract}

\maketitle

\tableofcontents

\section{INTRODUCTION}

The nature and condition of galaxies is predominantly determined by studying the light emitted by their stellar content. This has been true since the beginning of their study when \citet{1929PNAS...15..168H} determined the distance to M31 using the properties of individual stars, thus revealing the true nature of galaxies as other “island universes” like our own Milky Way Galaxy. Since this time our understanding of galaxies has evolved and expanded, and its interpretation has reached a level of complexity that requires a detailed understanding of stellar structure and evolution, stellar atmospheres and star formation. 

The theory of stellar structure and evolution allows us to calculate the luminosity, surface temperatures and gravities of stars with different masses and composition, and the length of time they exist with these parameters. The modelling of stellar atmospheres yields the electromagnetic spectrum emitted by stars with these 
surface parameters. Our understanding of star formation reveals the number of each of these stars we expect to be born by collapsing gas clouds. Combining these elements predicts the emission of light from a stellar population 
for comparison to that observed from galaxies and thus, in theory, reveal their ages, star formation histories and metallicities. This is particularly important for young stellar populations, whose light is dominated by massive ($M\gtrsim 10$\,M$_\odot$) stars. 

It is in the most massive, most luminous but shortest-lived stars that small changes in the assumed physics of stellar models can have the largest impact on the surface property of the stars, their impact on their surroundings and thus the interpretation of data. Their lives and deaths input tremendous amounts of radiative and kinetic energy into galactic environments. Not only do they have the highest luminosities of all stars and have strong stellar winds, but they are also the hottest stars, producing most of their light in the far-ultraviolet region of the electromagnetic spectrum. They emit substantial amounts of radiation shortwards of 912 angstroms, which ionizes surrounding hydrogen gas and powers the nebular emission of H\,II regions.  
In their deaths, when the core collapses to a neutron star or black hole, further energy and nucleosynthesis products are released in a supernova.  These likely pollute the environments of the first stellar halos with metals and provide the kinetic energy input to drive galaxy scale outflows. If the compact remnants retain an orbiting companion after the supernovae they can go on to accrete from this star and produce strong X-ray emission, again irradiating their host galaxies. At each stage of their lives and deaths, they produce observable signatures in the light of their stellar populations.

The earliest works attempting to synthesize stellar evolution, stellar atmosphere and star formation theory, with a focus on massive stars, into models for comparison with galaxy data are outlined by \citet{1968ApJ...151..547T}, \citet{1971ApJS...22..445S}, \citet{1971Ap&SS..12..394T} and others. 
The field took a significant step forward with the application of growing computational resources around the beginning of the 21st century. These permitted the creation of population and spectral synthesis codes such as Starburst99 \citep{1999ApJS..123....3L} and the GalaxEv models of \citet{2003MNRAS.344.1000B}. Here again the method of combining different strands of astrophysics was applied but with computers now able to take on the number crunching of detailed stellar structure models, rather than a reliance on analytic formalisms and simple models for evolution of stars on the Hertzsprung-Russell diagram. The detail and depth of such synthetic spectral models provided a crucial tool to reveal the history of the ever-larger samples of galaxies being revealed in the growing catalogue of observational data \citep[e.g. ][]{2004MNRAS.351.1151B}.

However, it has become apparent that these models, while extremely valuable, were incomplete. Unlike galaxies in the nearby Universe, galaxies at the highest redshifts were dominated by stellar populations that were not well described, 
with synthetic spectra unable to reproduce high ionization spectral features such as the He\,{\sc II} and C\,{\sc IV} stellar wind lines \citep{2003ApJ...588...65S}. As the observational frontier pushed into the first gigayear of galaxy evolution, it also highlighted the problem of where the ionizing photons responsible for reionizing the Universe at the end of the cosmic dark ages came from \citep[e.g.][]{2004MNRAS.355..374B,2004ApJ...610L...1S}. The extant models of the time struggled to produce sufficient hard radiation to permit galaxies to ionize their surroundings, without assuming extremely high ionizing photon escape fractions and steep luminosity functions. This presented a problem as other sources such as AGN were already known to have a low number density at the epoch of reionization \citep{2004ApJ...613..646D}.

Theorists soon rose to the challenge with two stellar-evolution-motivated solutions being put forward. First it was discovered that many massive stars experience rotation rapid enough to affect their evolution. This leads them to become hotter and more luminous than a non-rotating star, with extended main sequence lifetimes \citep[e.g., see][and references therein]{2000ARA&A..38..143M,2012A&A...537A.146E}. Secondly, it was found that interactions between stellar components in binary stars systems could significantly alter the evolution of massive stars \citep[e.g. see][and references therein]{2017PASA...34....1D,2017PASA...34...58E}. Again the primary effect of binary evolution was to produce more luminous and hotter stars. In both cases the result is to boost the ionizing photon flux inferred for a population based on its optical and ultraviolet photometry, and also make the far ultraviolet radiation harder. However both of these effects are relatively difficult to identify with the limited data available for distant galaxies, so determination of which was the correct interpretation was challenging. 

The deadlock was effectively broken by determination of the binary fraction of massive stars. \citet{2012Sci...337..444S} revealed that the vast majority of massive stars ($M>10$\,M$_\odot$) are in binary star systems. Later work  \citep[effectively synthesised in an analysis by ][]{2017ApJS..230...15M} confirmed this and suggested that the fraction of stars born in binary star systems was significant at all masses, although massive stars have the highest multiple fractions. Indeed the latest results suggest that the most massive stars are probably very commonly found in triple and higher order multiple systems \citep{2014ApJS..215...15S}. Neglecting the effects of multiplicity on the evolution of a stellar population is thus counter to the overwhelming majority of observational evidence, particularly where the light emitted from that population is dominated by massive stars.

The latest population and spectral synthesis models that use either theoretical or empirical initial binary parameter distributions produce a good match to observed stellar populations, providing strong evidence that binary stars must be accounted for in any model stellar populations \citep[e.g.][]{2016ApJ...826..159S}. Single star model populations that achieve the same results require the majority of stars to be born rapidly rotating, for which there is  limited evidence. However these effects are not independent. Mass and angular momentum transfer between stars in a binary can lead to boosted rotation and it is becoming apparent that the rotation of stars in binary systems must also be accounted for to fully explain stellar populations, especially at the highest redshifts. 

However while the current state-of-the-art population synthesis models have begun to more closely reproduce the observable properties of galaxies across cosmic time, there are still several places where a mismatch exists. Strong emission lines from extreme galaxies, for example, indicate that more high energy ionizing photons are produced than are currently predicted by these models \citep[e.g.][]{2019A&A...621A.105S}. 


This mixture of success in interpreting observational data and remaining challenges has also been complemented by a new source of constraints on stellar evolution theory. For example, asteroseismology is allowing us to peer within stellar interiors and constraining the size of convective cores during the main-sequence and determine the amount of mixing outside of the cores \citep{2021NatAs...5..715P,2021arXiv210709075J}. Binary compact remnant systems resulting from the evolution of massive star binaries can continue to evolve through gravitational wave radiation. The observation of these massive black hole binaries merging to produce astrophysical transients, detectable by gravitational wave observatories, has provoked a new interest in their progenitor systems and so gives a new window into the very early Universe \citep{2016PhRvL.116x1102A}.

These discoveries require a rethink of the ways in which the young and massive stellar populations, which dominate at early cosmic times and in some local environments, are modelled. This is where we begin this review of our current understanding of massive stars and their impact on our understanding of the galaxies they inhabit. We will summarize some of the current work on the physics of stars and their evolution, as well as reviewing empirical progress as the community considers what the latest and future observations of high redshift galaxies are starting to reveal about their stellar populations.

\section{THE CHALLENGE OF POPULATION SYNTHESIS}\label{sec:SPS}

In the very local Universe,
the impact of physics on the evolution and spectral contributions of individual massive stars can be identified through direct observation either of the stellar light, their death in supernovae or the remnants they leave behind.  At cosmological distances however, resolving individual stars is usually impossible, and instead analyses are based on the integrated light of entire stellar populations.

As a result it is necessary to combine state of the art models for individual massive stars with those for lower masses, and to match these models with calculated spectra which describe their expected emission. This process is known as population and spectral synthesis and has a long history dating back at least as far as the pioneering work of Beatrice Tinsley (\citeyear{1968ApJ...151..547T}). For more technical detail on the population synthesis process we refer readers to the relatively recent review by \citet{2013ARA&A..51..393C}. 

\begin{marginnote}
\entry{SPS}\hfill
Stellar population synthesis
\entry{SSP}\hfill
Simple stellar population in which all stars are the same age.
\entry{CSP}\hfill
Composite stellar population in which stars of different ages are combined according to an SFH
\entry{SFH}\hfill
Star formation history
\entry{SFR}\hfill
Star formation rate - this can be instantaneous (i.e. at time of observation) or averaged over some time period.
\entry{IMF}\hfill
Initial mass function (see section \ref{sec:imf})
\end{marginnote}

The model outputs must predict observable characteristics of an instantaneously formed stellar population as a function of age. However very few stellar populations in the Universe are formed from a single starburst with a well defined set of initial parameters. Thus most comparisons with data will also require combination of simple stellar populations according to some star formation history, over which the metallicity or other parameters in the component individual stellar populations may also vary \citep[see e.g.][]{2019ApJ...876....3L}. Where young stellar populations contribute significantly, the effects of nebular gas on the starlight may also need to be considered, requiring modelling of radiative transfer  through intervening material \citep[e.g.][]{2017RMxAA..53..385F}. Similarly dust absorption and reemission will modify the light from most galaxies. Dust extinction substantially affects the ultraviolet and blue optical light from galaxies, while thermal reemission is seen primarily in infrared and submillimetre  \citep[e.g.][]{2019A&A...623A.143F}. 

Thus most stellar population synthesis outputs must be combined with non-stellar components and a complex fitting operation 
undertaken when comparing with observational data. Increasingly this is done through application of Monte Carlo methodology to recover the posterior probability distribution associated with each input parameter 
\citep[e.g.][]{2018MNRAS.480.4379C, 2021ApJS..254...22J}, although this can be impractical where large catalogues of galaxies are considered or data on individual sources is sparse. Caution must be exercised when fitting photometry for faint or distant galaxies, as the number of free parameters being fitted can become larger than the number of data points unless additional assumptions are made. Spectroscopy will typically provide more constraints, but observations with narrow wavelength ranges can still be insensitive to some or all of the properties that shape a stellar population's spectral energy distribution.

The majority of widely-used population and spectral synthesis (SPS) codes \citep[e.g.][]{2003MNRAS.344.1000B,2009ApJ...699..486C} do not include prescriptions for binary evolution pathways. While this compromises their ability to predict young stellar populations dominated by massive stars, it has little effect in the mature, metal-rich environs of the local Universe. The Starburst99 \citep{1999ApJS..123....3L} model grid has recently incorporated a prescription for the inclusion of a limited range of binary evolutionary pathways \citep{2019A&A...629A.134G}, and also allows use of the Geneva library of rotating star atmosphere models \citep{2012A&A...537A.146E,2014ApJS..212...14L}. Binary spectral synthesis requires a significantly larger grid of stellar evolution models. Publicly released binary spectral synthesis models are limited to the Yunnan models of \citet{2015MNRAS.447L..21Z}, 
which are constructed using the BSE rapid stellar evolution code \citep{2002MNRAS.329..897H}, and the Binary Population and Spectral Synthesis code \citep[BPASS, ][available from \url{https://warwick.ac.uk/bpass} or \url{https://bpass.auckland.ac.nz/}]{2017PASA...34...58E, 2018MNRAS.479...75S}, which uses a custom grid of detailed binary evolution models. There are also the earlier models from the Brussels group which are not publicly available \citep[e.g.][]{2003A&A...400..429B}. 

\begin{summary}[SUMMARY POINTS]
\noindent A population and spectral synthesis requires four key ingredients:
\begin{enumerate}
\item  Stellar evolution models at sufficient mass resolution to capture behaviour at key ages and mass ranges. If modelling multiple systems, these must also span a full range of multiple parameters and track the binary evolution.
\item A parameterisation for the initial properties of the population, including composition (metallicity, relative elemental abundances), distribution in mass (IMF), rotation (initial velocity, angular momentum loss or gain) and multiplicity parameters (initial period, multiple fraction, initial mass ratios) where appropriate.
\item Electromagnetic spectra as predicted from stellar atmosphere models, as a function of surface temperature, surface gravity, composition and assumed stellar wind parameters, at sufficient resolution to compare with observations.
\item Prescriptions for supernovae, remnants and the evolution of multiples containing remnants.
\end{enumerate}

\noindent It must generate one or more of the following outputs:
\begin{enumerate}
    \item The integrated light electromagnetic spectrum of the modelled simple stellar population as a function of age.
    \item (optional) The number counts and total masses of living stars at each timestep.
    \item (optional) Supernova and other transient event rates.
    \item (optional) Reprocessing of starlight by dust or nebular gas.
\end{enumerate}

\end{summary}

\begin{marginnote}[]
For illustrative purposes, we will refer to a fiducial stellar population model:
\entry{SPS Model}\hfill
{BPASS} v2.2.1

\entry{Star Formation History}\hfill
SSP

\entry{IMF}\hfill
Broken power-law \hfill
$\alpha=-1.30$ for $0.1<$M$<0.5$\,M$_\odot$.
$\alpha=-2.35$ for $0.5<$M$<300$\,M$_\odot$.
\end{marginnote}

\section{RECENT DEVELOPMENTS IN STELLAR EVOLUTION THEORY}
\label{sec:stellar}

\subsection{The initial conditions}

\subsubsection{The initial mass function}\label{sec:imf}

A significant factor in the impact of stars on galaxies is the relative fractions of different types of stars in any given stellar populations. The initial conditions for star formation, predominantly manifest through the initial mass function (IMF), are key to estimating the number of stars in a population as a function of mass, and thus calculating stellar contributions to the luminous, mechanical and nucleosynthetic outputs to a galaxy. 

Measuring the IMF of a stellar population is difficult and is dependent on the mass estimates of individual stars, which have varying accuracy \citep[e.g.][]{2021A&ARv..29....4S}. However it appears that the stellar IMF, to a good approximation, is well represented by a power law distribution in the range 1-100\,M$_\odot$ with massive stars rarer than their low mass counterparts, and with little evidence for variation over the history of the Universe. The currently assumed power law slope for most stellar populations is still surprisingly similar to the very first estimate proposed by \citet{1955ApJ...121..161S}, which takes the form $N(M)\propto M^\alpha$ with $\alpha=-2.35$. However growing evidence suggests that there is indeed variation in the inferred IMF power law slope from galaxy to galaxy and from stellar cluster to stellar cluster \citep[see e.g.][]{2018PASA...35...39H}. 
It has also become clear that the IMF deviates from a simple power law at low stellar masses \citep[e.g.][]{2003PASP..115..763C}. Accurately constraining both ends of the distribution is essential for translating observed light into an estimate of mass, star formation rate and related parameters (see Section \ref{sec:SFRs}).

However, as we discuss below it is the most massive stars, essentially above $\approx 10 M_{\odot}$ that have the greatest impact on their surrounding environment. Thus the IMF slope above this mass and the maximum mass for a star 
are important parameters that affect predictions of spectral synthesis. The upper mass for stars was initially thought to be around 100~M$_{\odot}$ from observations in our Galaxy \citep{2005Natur.434..192F}. However observations of the Large Magellanic Cloud have indicated that stars as massive as 300~M$_{\odot}$ can form \citep{2010MNRAS.408..731C}. These 
stars produce 30 to 40\% of the ionizing photons for the region they inhabit, despite representing a much smaller fraction of the mass. 
Detailed analysis of the same region has indicated that the IMF slope for massive stars could be significantly shallower than Salpeter \citep{2018Sci...359...69S} indicating there are many more massive stars than might be expected. 

\begin{marginnote}
\entry{Commonly used stellar IMFs} \

\entry{Salpeter (1955)}\hfill
$N(M)\propto M^\alpha$, $\alpha=-2.35$ for $0.1<$M$<100$\,M$_\odot$
\newline


\entry{Kroupa (2001)}\hfill
$N(M)\propto M^\alpha$, 
$\alpha=-0.3$ for M$<0.08$\,M$_\odot$,

\smallskip

$\alpha=-1.3$ for $0.08<$M$<0.5$\,M$_\odot$,

\smallskip

$\alpha=-2.3$ for $0.5<$M$<120$\,M$_\odot$,
\newline

\entry{Chabrier (2003)}\hfill
$N(\log M) \propto e^{- (\log M - \log 0.08)^2}$
for $0.08<$M$<1$\,M$_\odot$,

\smallskip

$N(M)\propto M^\alpha$ with 
$\alpha=-2.3$ for M$>1$\,M$_\odot$,

\end{marginnote}

One problem with the IMF is how to implement it when creating a synthetic stellar population. Typically functional forms involving either broken power-laws \citep[e.g.][]{1993MNRAS.262..545K} or Schecter functions \citep[e.g.][]{2003PASP..115..763C} to flatten the low mass end of the distribution are used to compute the number of each stellar model mass within a stellar population formed with a certain mass of star formation. While this statistical sampling is reasonable for large SFRs, where lower masses of stars are formed stochastic sampling of the IMF becomes important to consider \citep[e.g.][and reference therein]{2015MNRAS.452.1447K}.

A final complication is that it is typically assumed that the IMF is universal and constant across cosmic time. As discussed above, there is variation even locally, and in the distant Universe a completely different regime may dominate. Simulations indicate that at least the first stars in the Universe (known as Population III and formed from gas enriched only by Big Bang nucleosynthesis) had an IMF that was very top heavy compared to that for later stellar generations \citep{2004ARA&A..42...79B}. Exactly when the significant change to metal-dominated stellar envelope opacities happened is uncertain, although recent work by \citet{2021arXiv210708634S} suggests a metallicity threshold $Z=10^{-5}$ to $10^{-4}$. This is low enough that only small fraction of stars formed in the Universe would meet this criterion. Current observations suggest that the limit is probably below the lowest metallicities observed to date for high redshift galaxies \citep{2020ApJ...903..150J}, although extreme emission line galaxies show hard spectra that may hint at very low metallicity populations (see Section \ref{sec:spec}). 

\subsubsection{Initial stellar rotation and binary parameters} \label{sec:params}

If rotation or duplicity are present in a stellar population,  then extra initial parameters are required. 

The simplest parameter sets the number of primary stars at each mass that are in binaries: the mass-dependant binary fraction. We also require an initial period distribution, initial eccentricity and the initial mass ratio of the two stars, these are also typically mass-dependant, and the mass ratio distribution may further depend on the period. We note that \citet{2002MNRAS.329..897H} determined that the period and eccentricity can be combined into a semi-latus rectum distribution to reduce the number of initial parameters required. Other factors that can be added if more complexity is required include the initial spins and orientation of the two stars. Thus the initial assumptions going into creating synthetic stellar populations begin to grow in number \citep[see][for further details]{2017ApJS..230...15M}. Significant advances have been made in showing that the most massive stars are nearly all in binary star systems and accurately determining the inital binary fraction, period distribution and mass ratios \citep{2012Sci...337..444S,2019ApJ...875...61M}. These are currently implemented in the detailed binary population spectral synthesis code BPASS, the models of \citet{2019A&A...629A.134G} and several population synthesis codes but not in the other spectral synthesis codes that include interacting binary stars.

Whether these stars are born spinning and retain their angular momentum for a substantial fraction of their lifetime will also affect their evolution. As a result eccentricity and spin are approximated in addition to binary parameters in rapid population synthesis codes which typically focus on stellar numbers rather than detailed spectra. Our knowledge of the rotation rates of stars when they enter the main sequence is incomplete, especially for massive stars given that so many are in binary or multiple star systems. But it is likely that few stars are born rapidly rotating and most begin only with a moderate rotation, generally spinning down over their evolution due to mass-loss removing angular momentum \citep{2013ApJ...764..166D}. It is important to note that measurements (i.e. $v \sin i$) only hint at surface rotation, while the internal rotation frequency profile has a larger effect. Current knowledge on internal rotation across stellar evolution is summarised by \citet{2019ARA&A..57...35A}. 

\subsubsection{Initial composition}

After its initial mass, the metallicity of a star is the next most important factor in how it will evolve throughout its life. The composition affects a star's evolution through the strength of stellar winds and the catalysis of hydrogen fusion. 
Of primary importance is iron, which makes a large contribution to opacity in stellar interiors. This sets the strength of the mass-loss of radiatively-driven winds \citep[e.g.][]{2001A&A...369..574V}. Its abundant electron transitions generate many weak absorption lines, making stellar material opaque and thus able to interact with the radiation field to accelerate material away from the surface. To a lesser extent this opacity also causes stellar radius to increase with iron abundance, again due to radiation pressure. This slightly lowers the surface gravity of stars at higher metallicities.

The carbon-nitrogen-oxygen (CNO) elements have a secondary impact, but are still significant due to their catalysis of hydrogen fusion. Although varying the CO abundances does have some impact on stellar structure, it is of minor importance compared to the significant impact of the changing the iron abundance. The amount of CNO elements has a bigger impact on the conditions of hydrogen fusion; when less CNO is available a stellar core has to become denser and hotter to support itself against collapse. 

\begin{marginnote}
\entry{Z} \hfill Metallicity mass fraction, comprised by elements heavier than helium.
\entry{Z$_\odot$} \hfill Solar metallicity, often defined as Z$_\odot$=0.020, although current observations favour Z$_\odot$=0.014 \citep{2021A&A...653A.141A} 
\entry{12+log(O/H)}\hfill  Oxygen abundance in a form traditionally measured by observers. Solar metallicity is usually defined as 12+log(O/H)=8.7 \citep{2021A&A...653A.141A}.
\entry{[Fe/H]} \hfill  The logarithm of the iron abundance ratio by number (rather than mass), scaled relative to the Solar value, i.e. $\log_{10}\left(\frac{\mathrm{Fe/H}}{\mathrm{(Fe/H)}_\odot}\right)$.

\end{marginnote}

A problem arises here in that the most straightforward observational signatures of metallicity are the oxygen abundance, which is then often used to estimate the total metallicity of the stars - particularly at high redshift where signal to noise is limited \citep[e.g.][]{2016ApJ...826..159S}. However these early galaxies typically have a composition which is significantly oxygen-rich and iron-poor \citep{2016ApJ...826..159S,2019MNRAS.487.2038C} compared to the Solar reference composition. Thus discussing a bulk metallicity often lacks the subtlety required for detailed interpretation. We should consider the iron abundance in selecting which stellar population models we should be using, and  oxygen abundance to further refine our models.

\subsection{Mass-loss rates in stellar winds}\label{sec:winds}

All stars lose mass in a stellar wind driven either by metal line opacity or by the balance between radiation pressure and gravitational attraction at their surface. In general the more luminous a star, the stronger that wind. 
While, on the main sequence, mass loss is moderate, the strength of mass loss increases thereafter, whether the star evolves to become a cool red supergiant or a hot Wolf-Rayet (WR) star. For the former the low surface gravity accelerates the rate at which mass can be lost. For WR stars, increased mass-loss arises from a combination of high luminosity-to-mass ratios, which result from prior mass loss reducing a star to its helium core, and an iron-opacity bump at a temperature of $10^5$\,K causing significant opacity near the surface layers of these stars. 
This means that winds are also dependent on the initial iron abundance of the star. Lower metallicity stars like those found in early galaxies typically have weaker stellar winds.

As a star loses mass it becomes less luminous and less massive. Thus winds can slightly increase stellar lifetimes, especially if a significant fraction of the initial mass is lost. How this changes a star's impact on its surroundings is complex. Stars with weaker winds provide less kinetic energy input to their environs, but will shine with higher luminosity, and for a shorter overall lifetime. The only way this behaviour can be captured  is to create detailed stellar evolution models and include them in a full synthetic population model.

To take account of mass loss in a stellar model, a recipe is required to provide the rates as a function of stellar parameters. The currently widely use formalism is a mass-loss prescription that uses \citet{2001A&A...369..574V} for OB stars, \citet{2000A&A...360..227N} for WR stars and \citet{1988A&AS...72..259D} for cool supergiants and all other stellar types. These match observed stellar population reasonably well \citep[e.g.][]{2017PASA...34...58E}. However, there is significant uncertainty over the exact validity of this simple scheme and recent works improve on these predictions for OB stars \citep[e.g.][]{2021A&A...648A..36B,2021MNRAS.504.2051V}, red supergiants \citep[e.g.][]{2020MNRAS.492.5994B} and WR stars \citep[e.g.][]{2020MNRAS.499..873S}. There are also hints that the O-star mass-loss rates for hot main sequence stars may need a slight revision \citep{2021arXiv210808340H}. 
Each of these studies have compared theoretical mass-loss rates against observed stars, finding modest improvements in fitting the data. This area is discussed in greater detail in the review by Vink in this volume. However, revised wind rates have yet to be included in any stellar population model of which we are aware. The impacts of revised rates may be significant in some regimes but evidence suggests that 
their overall impact on predicted massive star populations will be limited. For example, while mass-loss rates will impact on the relative number of WR stars and red supergiants observed in a population, the ratio in number of these stars in nearby galaxies, and its dependence on metallicity, is actually already well produced with extant binary population synthesis models \citep{2021arXiv210708304M}. 

\begin{marginnote}
\entry{Metallicity dependence of winds} \hfill Mass loss rates vary with metallicity. Typically $\dot{M} = \dot{M_{\odot}} Z^{\alpha}$ is assumed, with $\alpha$ ranging from 0.5 to 0.86 \citep[e.g.][and the chapter by Vink in this volume]{2001A&A...369..574V}. Since stellar winds are driven by the opacity of iron-lines in the stellar atmosphere, fewer metals implies lower opacities and weaker winds. The exponent tends to one at very low metallicity and decreases at higher metallicities as the lines become saturated.

\end{marginnote}

The optically-thick, high velocity winds driven by Wolf-Rayet stars  are amongst the most uncertain. Modelling such stellar atmospheres and winds is complex \citep[e.g.][]{2020MNRAS.499..873S,2021MNRAS.503.2726H} and the stellar surface cannot itself be observed. However theoretical and observational progress is being made, with indications that these winds are strongly metallicity dependent and that they also depend on surface composition (resulting from dredged up material rather than initial abundances) as well as luminosity. A key unknown is how the high mass-loss rates of WR stars compare with rates at the lower masses of similar stripped-envelope stars made in binary systems. Relatively few such systems have been identified to date \citep[e.g.][]{2008A&A...485..245G}. Current theoretical work suggests extrapolation from WR star mass loss rates may significantly over-predict the mass loss rates of these stars \citep{2017MNRAS.470.3970Y}.

However the largest current unknown regarding mass-loss rates is what happens to the most massive and most luminous stars. 
If the mass-loss rates are high enough, they may limit the maximum mass of stars that can be observed within a stellar population, with  a strong effect on the ionizing and kinetic energy output of a stellar population. How the mass loss of the most massive stars relates to the enigmatic luminous-blue variable (LBV) stars also remains unclear. Originally it was thought that all LBVs must be very massive stars, although the picture is still unclear \citep[e.g][]{2018AJ....156..294A}.

\subsection{Rotation}\label{sec:rotation}

\begin{figure}
\includegraphics[width=0.9\textwidth]{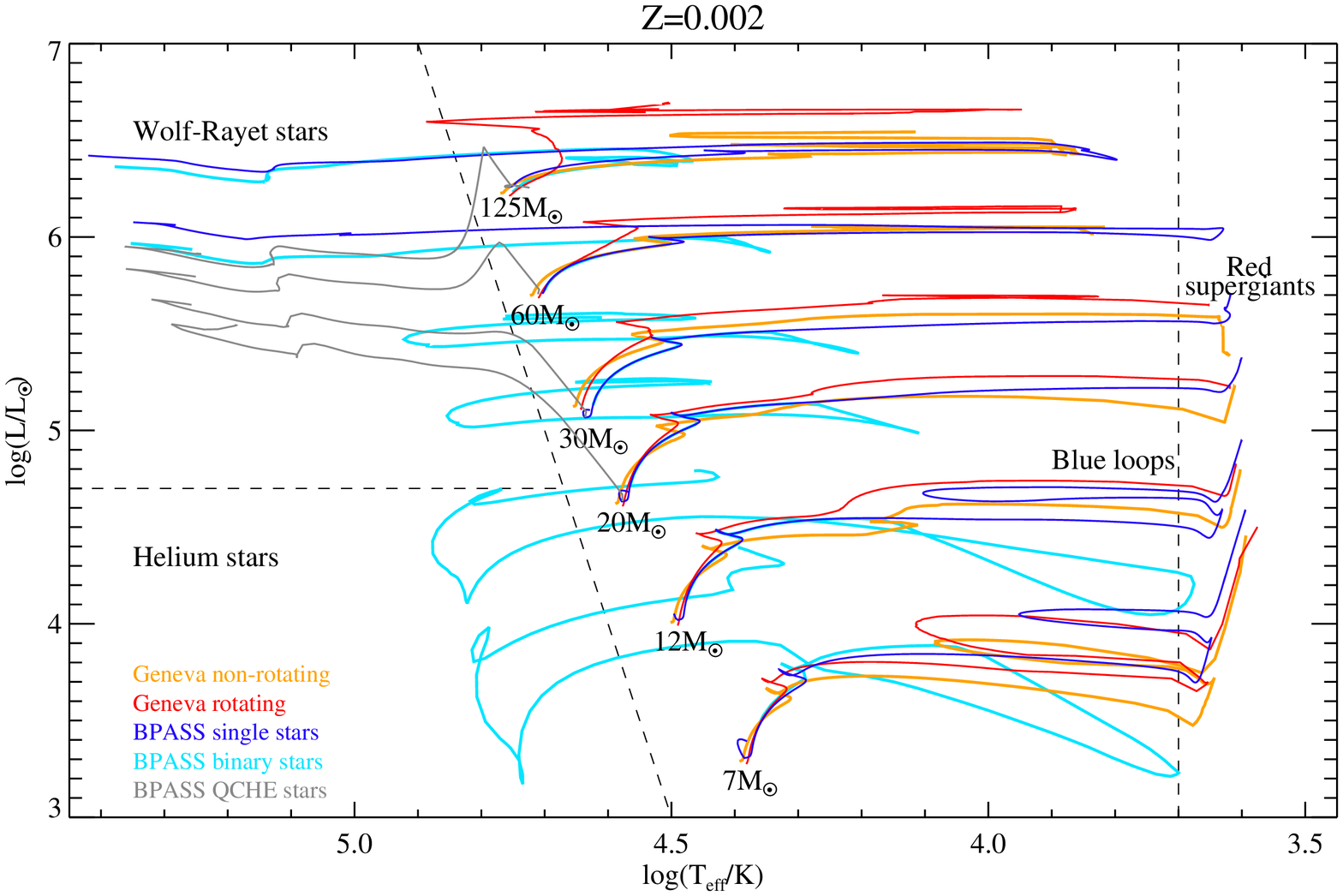}

\includegraphics[width=0.9\textwidth]{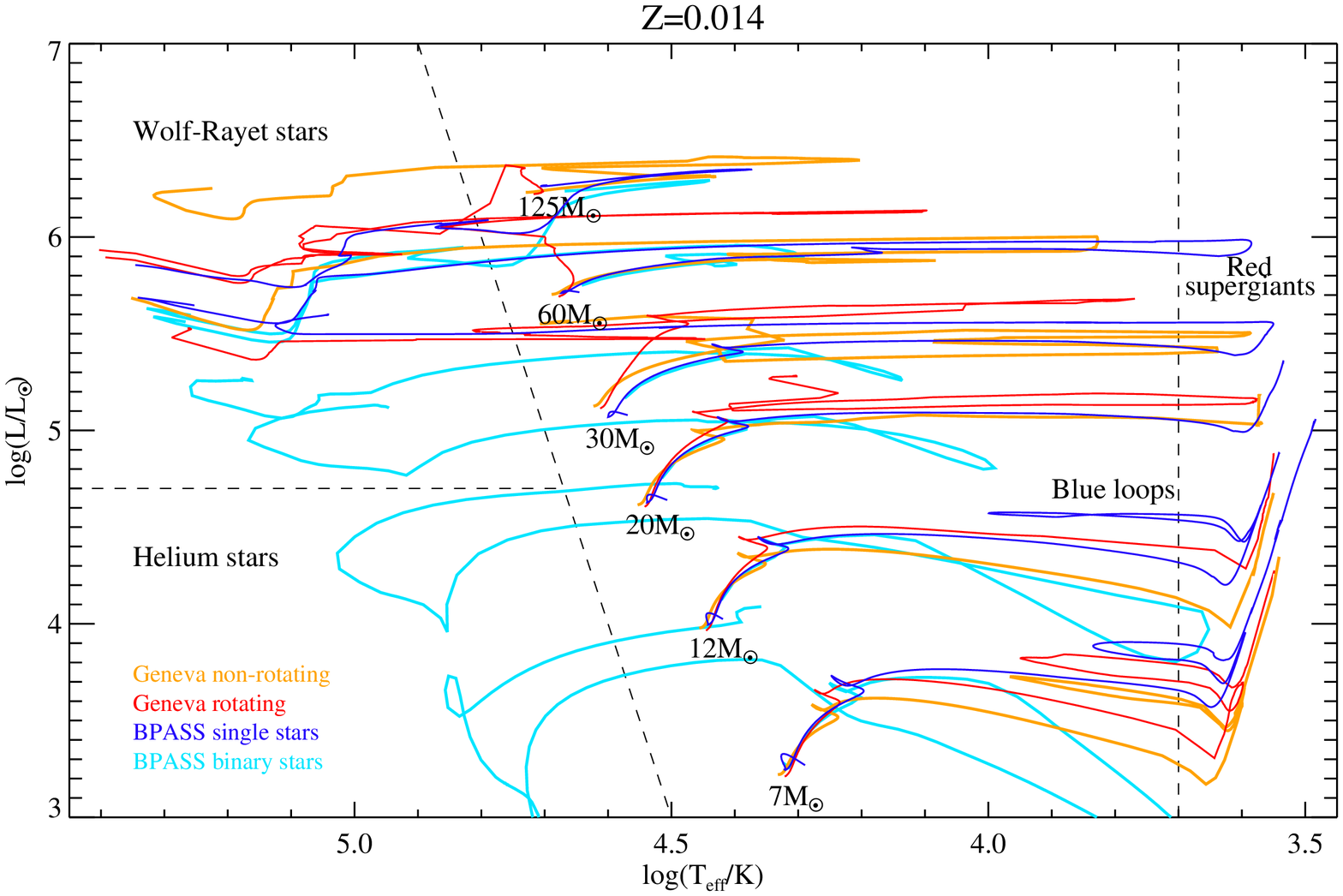}
\caption{Example Hertzsprung-Russell diagrams of BPASS \citep{2018MNRAS.479...75S} and Geneva \citep{2012A&A...537A.146E,2013A&A...558A.103G} stellar evolution models. Non-rotating Geneva models are shown in orange and rotating models in red in increasing luminosity the masses are 7, 12, 20, 32, 60 and 120~M$_{\odot}$. Blue lines show BPASS single-star models and the light-blue lines are the BPASS binary tracks with primary stars that have initial masses of 7, 12, 20, 30, 60 and 125~M$_{\odot}$. In the $Z=0.002$ panel we include BPASS quasi-chemically homogenously evolving stars with the greyk lines at masses of 20, 30, 60 and 120~M$_{\odot}$.}
\label{fig-hr-single}
\end{figure}

\begin{marginnote}
In a stellar interior it is typically assumed that mixing only occurs within convective zones. Within radiative zones mixing is not assumed to occur although there are many possible mechanisms that occur to provide mixing of some degree \citep{2021NatAs...5..715P,2021arXiv210709075J}
\end{marginnote}

The simplest model of a star assumes it to be single and non-rotating. For such stars, assumptions such as hydrostatic and local thermodynamic equilibrium can be made during evolutionary phases driven by the early nuclear fusion reactions of hydrogen and helium burning. However when the star rotates, the hydrostatic balance is altered by a centrifugal force, causing the equator to bulge outwards. This upsets the local thermodynamic equilibrium and so Eddington-Sweet currents are induced, producing extra mixing through radiative zones in the stellar interior, 
and so nuclear processed and unprocessed material can be mixed out of and into the stellar core. This alters surface abundances to become nitrogen rich and mixes fresh hydrogen into the core to extend the stellar lifetime \citep[e.g.][]{2012A&A...537A.146E,2016ApJ...823..102C}.
While this simple concept has been known for some time, more recent work has investigated in how rotation interacts with magnetic fields and how differential rotation can further enhance the strength of mixing. The physics is exceedingly complex \citep[e.g.][]{2005A&A...440.1041M,2012MNRAS.424.2358P}, especially considering that this inherently two-dimensional object is usually implemented within a one-dimensional stellar evolution code. Two-dimensional stellar evolution codes \citep[e.g.][]{2016JCoPh.318..277R} are rare and relatively slow, making construction of large model grids difficult.

\begin{marginnote}
Rotation generally makes massive stars longer lived, making a stellar population bluer at fixed age.
\end{marginnote}

The detailed interaction between rotation, magnetic fields and a star's evolution is complex. Evolution of a star naturally sets up a rotation profile within it. The core becomes increasingly dense as its mean molecular weight increases. This in turn may spin up the core due to conservation of angular momentum, although this is uncertain. At the same time the envelope expands, slowing down for the same reason, with further braking of rotation being caused by any stellar winds. Recent models suggest that, while massive stars arrive on the main sequence with a range of rotation rates, mass loss and magnetic breaking drives most rapidly towards a low surface rotation rate \citep[e.g.][]{2020MNRAS.493..518K}. How the core links to the envelope is uncertain, but if differential rotation can drive a strong magnetic dynamo then this may force solid body rotation upon the star.  However, only by combining asteroseismology measurements with detailed modelling will understanding be achieved \citep{2019ARA&A..57...35A,2019MNRAS.485.3661F,2019A&A...631L...6E}.
We note that mass loss from a star can also be enhanced due to rapid rotation of a star. This in turn moderates the rotation rate for a star with any reasonable stellar winds. The exceptions to this will be at the lowest metallicities where mass loss is already weak  \citep{2005A&A...443..643Y}. At low metallicities, stars may remain rotating rapidly for much of their evolution and experience very different evolution by being fully mixed on the main sequence, becoming WR stars without first evolving through the red supergiant phase.


\begin{marginnote}
Another rotation-induced evolution pathway, but this time only seen in multiple systems is chemically quasi-homogeneous evolution (see section \ref{sec:qche}).
\end{marginnote}

\begin{figure}
\includegraphics[width=\textwidth]{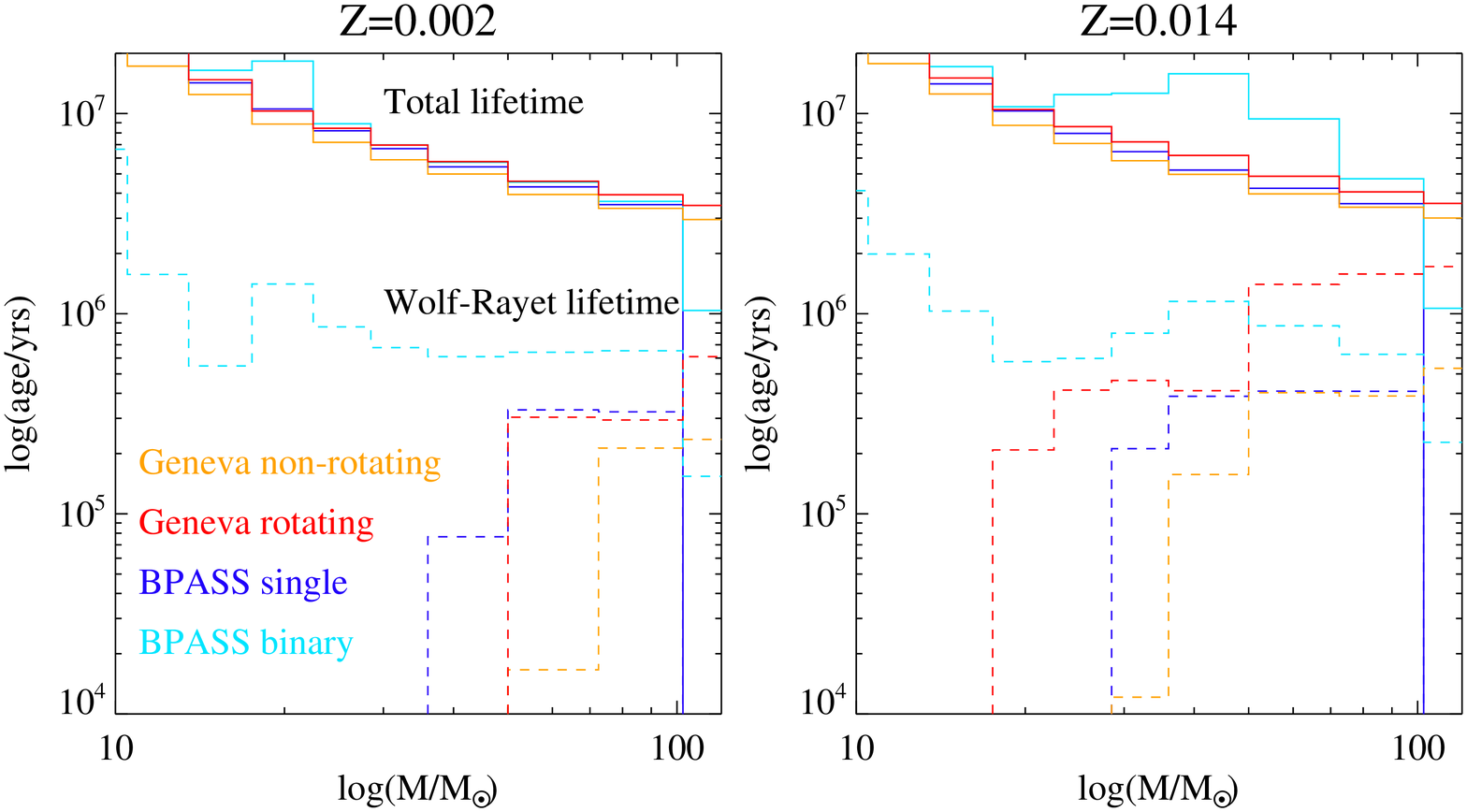}

\caption{Mean lifetimes as a function of mass for stellar evolution models drawn from BPASS and Geneva grids at two different metallicities. Solid lines are for total lifetime and dashed lines for Wolf-Rayet/Helium star lifetime. Colours are the same as in Figure \ref{fig-hr-single}.}
\label{fig-lifetimes}
\end{figure}

We illustrate the impact of including rotation in stellar evolution models in Figure \ref{fig-hr-single}. Here we compare models drawn from the Geneva stellar evolution grid \citep{2012A&A...537A.146E,2013A&A...558A.103G}, with and without rotation.  Models which include rotation (an initial rotation rate of 0.4 of the critical break-up velocity) have extended main sequences and also weaker blue loops during helium burning. Both of these are direct results of rotational mixing. As Figure \ref{fig-lifetimes} shows, rotation also extends the total lifetime of the star and particularly increases the WR star lifetime. However few stars become WR stars that wouldn't have done so in non-rotating models. 

We note that the latest results from asteroseismology suggest that the mixing induced by rotation may in fact be more variable between stars. In addition there may be other extra mixing mechanisms beyond those included in the models presented in Figure  \ref{fig-lifetimes} as found by \citet{2021NatAs...5..715P} and \citet{2021arXiv210709075J}. However these observations only include stars with $M<24\,M_{\odot}$, while the most important stars for young stellar populations are more massive than this limit and the mixing profiles of such stars are unknown. 

\subsection{Binary and Multiple Stars}\label{sec:binaries}

\subsubsection{The road to understanding the importance of binaries} 

Historically, the study of binary and multiple stars 
and their importance in stellar populations of massive stars has been largely ignored, even though there has always been strong evidence that many stars are in binary or multiple systems.  
It was known early on that WR stars were helium stars, formed by loss of the hydrogen layers from massive stars. The exact mechanism of mass loss was uncertain, although \citet{1967AcA....17..355P} did initially suggest binary interactions as a possibility. However then stellar winds were investigated 
and suggested that massive stars could lose their hydrogen envelopes through winds alone.

A growing recognition of the importance of binary evolution began early in the 21st century, partly inspired by evidence from an unexpected direction.  
Ultraviolet spectra of galaxies at $z\sim3$ \citep{2003ApJ...588...65S} showed stellar emission lines from O stars and WR stars in numbers that extant single star models were unable to explain. 
Studies including binary interactions in massive binary stars had already begun to accurately map out the many differences to single star evolution \citep[see][for a detailed review of modelling of binary stars]{2017PASA...34....1D}. At this time BPASS binary evolution models were first created in order to explain resolved stellar populations and the relative rate of different supernovae binary interactions \citep{2008MNRAS.384.1109E}. 
To apply such models to unresolved stellar populations required 
creation of a binary star spectral synthesis which required extensive additional physics and many extra stellar models. 
Tests on nearby galaxies \citep{2009MNRAS.400.1019E} were followed by analysis of the distant Universe, and this demonstrated that including binary interactions and simple rotational prescriptions within these systems could explain the \citet{2003ApJ...588...65S} observations \citep{2012MNRAS.419..479E}. 

The fact that binary stars are important for understanding early galaxies was initially far from widely accepted, but was supported by other evidence that began to appear around the same time. This includes observations of the extremely high initial binary fraction amongst massive stars within our Galaxy \citep{2012Sci...337..444S}, observations for the progenitors of type Ib/c supernova suggesting binary interactions have a significant role in determining the fate of this population of stars  \citep{2012A&A...544L..11Y,2013MNRAS.436..774E}, and most recently the detection of merging black hole binaries in gravitational wave transients that are mostly (but not only) arising from isolated binary evolution \citep{2016PhRvL.116x1102A}. However even today many studies continue to use models which neglect binary interactions, and the significance of their impact in different regimes is still debated.

\subsubsection{Differences introduced by binary interactions}\label{sec:diffs}

It is important to ask why binary populations are so important and different from the products of single star evolution. The key difference is that no longer are the initial mass, spin and composition of a star the primary features that determines their future evolution. Stars can interact with their companion and mass transfer can occur between them. The initially more massive star evolves relatively faster and can lose mass by transfer to its companion. Angular momentum can also be transferred in this process or through tidal interactions between the two stars if they are sufficiently close.

A key difference between the mass loss from stellar winds and that in a binary system is timescale. Binary mass transfer occurs when a star grows in size such that it fills its Roche lobe. This is the equipotential surface beyond which material is no longer gravitationally bound to the star, due to the combination of the companion star's gravity and centrifugal forces of the binary orbit. If this occurs on the main sequence, mass transfer is typically stable as a star's structure can adjust to remain just within the Roche lobe. If mass transfer occurs post-main sequence, for example as a star crosses the Hertzsprung gap, mass transfer tends to be unstable. The mass loss does not slow the increase of the donor star's radius and  can quickly increase in magnitude, such that a significant amount of mass is lost in a short time. In the most severe cases the donor star can continue to swell in size until it engulfs the companion star to enter common envelope evolution. In such events the majority of hydrogen envelope is lost within a period of ranging from a few years to a few thousand years, causing the binary orbit to shrink. These events can also lead to the two stars merging, resulting in unusual stellar structures.

Whatever the strength and stability of the mass transfer, the eventual result is similar. Once the donor star loses its hydrogen envelope it becomes a WR star or helium star, dependent on its eventual mass. Such stripping exposes the cores that result from main-sequence hydrogen fusion. WR stars are recognised observationally by their high luminosities and broad emission lines which are the result of strong, optically-thick stellar winds, and have masses down to around 5\,M$_{\odot}$. Below this mass the helium cores are no longer able to drive  optically thick winds, and instead are described as helium stars. While they fuse helium to carbon and oxygen, they remain compact with effective temperatures of $10^5$~K. There are few known examples in the mass range from approximately 1.5 to 5~M$_{\odot}$ (below this mass they are observed as sdOB stars), with the strongest candidate being HD\,45166 \citep{2008A&A...485..245G}. They are difficult to detect because in formation they will have transferred a significant amount of material to their companion star. This will be an O or B star, and being cooler this will outshine the helium star at optical wavelengths \citep[although the hot helium star may be brighter in the far ultraviolet,][]{2018A&A...615A..78G}.  
Such stars later evolve into helium giants and these stars dominate the population of type Ib supernova progenitors \citep{2012A&A...544L..11Y,2013MNRAS.436..774E,2016MNRAS.461L.117E,2017MNRAS.470.3970Y}. These stars can be seen in Figure \ref{fig-hr-single} in the light blue tracks which are lower luminosity than the WR stars created by the single star models.

A recent complication is the realisation that the luminosity and mass thresholds at which stars change observationally from helium stars to WR stars may be metallicity dependent \citep{2020A&A...634A..79S}. Lower iron abudance leads to lower opacity and hence is is more difficult to drive the optically thick winds required as WR observational signatures. Hence helium stars might be easier to first detect in lower metallicity environments.

While Wolf-Rayet stars can be formed by single star evolution through stellar wind mass loss, helium dwarfs can only be made by binary interactions. 
These lower mass helium stars only have weak winds,
unlikely to have been strong enough to remove their hydrogen envelope alone. However, being so hot, these stars output most of their light in the Lyman continuum and thus are important contributors to the ionizing budget of a stellar population, especially at ages $\sim$10-50\,Myrs when  the O stars in a stellar population have died and white dwarfs are yet to be born.  
The companion star in such systems will have accreted a significant amount of mass and, in some cases, angular momentum.  These too may therefore experience a different future evolution to that expected of a single star \citep{2007A&A...465L..29C}, with a shortened lifetime (although still longer than the main sequence lifetime of the primary). In extreme cases the companion is spun up, undergoing homogenous evolution (see sidebar).

\begin{textbox}[h]\section{Sidebar: Quasi-Chemically Homogeneous Evolution}\label{sec:qche}
The impact of rotation on evolution in single stars is uniquely determined by the initial spin of the star, as well as its metallicity and initial mass. A single rotating star cannot suddenly gain extra angular momentum. The only way for angular momentum to increase is for a star to be a member of a binary or multiple star system.
If a star accretes a substantial fraction of its initial mass from a binary companion:

\vspace*{10pt}

\begin{minipage}{1.1\textwidth}
\begin{itemize}
 \item The extra mass makes the star more luminous and shortens its lifespan. In some cases the initially less massive star may even experience the first supernova in the binary system. 
\item The additional angular momentum may also lead to the stars rotating fast enough that they become fully mixed due to rotationally-induced mixing.  
\item Quasi-chemically homogeneous evolution (QCHE) occurs at low metallicities, where such a star cannot lose the gained angular momentum and so remains rapidly rotating and hence fully mixed for the majority of its main sequence lifetime \citep{2006A&A...460..199Y}. Evolution tracks showing such behaviour are shown in Figure \ref{fig-hr-single}.
\item If rotation remains high until the end of evolution, the star may generate a long-gamma-ray burst, or possibly a superluminous supernova \citep{2007A&A...465L..29C}. In both cases it is  rapid rotation and its interaction with the forming black hole or neutron star respectively that drives the evolution of the transient. 
\end{itemize}
\end{minipage} 

\vspace*{8pt}

\noindent A second binary pathway also leads to QCHE. If the stars are born in a sufficiently tight orbit, tidal forces can be strong enough to cause both stars in the binary to experience this evolution \citep{2016A&A...588A..50M,2016MNRAS.458.2634M}. Whether both these pathways occur, or which dominates, is unknown, although indirect evidence from long-GRBs and GW transients requires that such systems must exist.
\end{textbox}

%

A final consequence to consider from binary interactions is the merger of the two stars. When a common envelope interaction occurs the hydrogen envelope is usually lost by transfer of the orbital binding energy into the envelope. In some cases there is not enough orbital energy to remove the entire envelope and so the two stars eventually merge. Such merger products can appear to be quite normal stars based on their surface properties. They may however also be rapidly rotating giants or have a significantly depleted hydrogen envelope due to the mass lost before the eventual merger. Merger products, with the range of possible interior structures, can exist across the HR diagram \citep[e.g.][]{2020MNRAS.496.5503B,2020MNRAS.495.2796S}. The best known example to date was the progenitor of supernova 1987A \citep[e.g.][]{2019MNRAS.482..438M}.

\subsubsection{X-ray binaries}\label{sec:XRBs}

An important phase of massive binary evolution occurs after the first supernova. Most binary systems will become unbound due to the combination of losing a significant amount of mass and the momentum kick received by the compact remnant in the core-collapse.  However if mass transfer has been efficient so the companion star is relatively massive, or if the supernova kick is in a specific direction, the system can remain bound. 
In close binaries, the companion star will eventually fill its Roche lobe and some of its mass will be transferred to the black hole or neutron star that it now orbits. The resulting hot accretion disk will be observed as an X-ray binary (XRB) during accretion episodes, but more challenging to observe in quiescence \citep{1993ARA&A..31...93V,2006ARA&A..44..323F}. Many star forming galaxies have significant XRB populations, and the total X-ray luminosity can be used as a tracer of moderate aged stellar populations. Detailed study of X-ray binaries and their number is starting to be included within spectral synthesis models \citep[][]{2000A&A...358..462V,2021arXiv210812438S}.

\subsubsection{Gravitational wave transients}

In massive systems, the second star will also undergo core-collapse, 
leading to the formation of a binary compact remnant system. The further impact on their host galaxies will be minimal but the eventual merger of such double compact objects provides a novel probe into  early stellar populations. The substantial evolutionary timescales of these systems require the progenitors of many of today's transient detections to have formed in the distant Universe.

The  study of gravitational wave transients is currently one of the fastest expanding fields of astrophysics. While they had been predicted for just over a century, the first detection in 2015 \citep{2016PhRvX...6d1015A} has driven significant interest in these events. 
A full review of this area is well beyond the scope of this article. However recent results coming from the study of these objects are having significant impact on our understanding of stellar evolution; first the fact that extremely massive black holes can form indicates that stellar winds must become weaker at lower metallicity environments \citep{2016PhRvL.116x1102A}; second that massive binary stars must be formed and numerous in early galaxies \citep{2021arXiv210714239M,2020PhRvL.125j1102A}; and third that X-ray binaries may boost the mass of some black holes and potentially produces strong ionizing radiation \citep{2020ApJ...897..100V}.
The fact there is a sharp upper mass limit to the observed black hole masses observed suggests an additional insight; that pair-instability and pulsation pair-instability supernovae occur. These events cause significant mass loss from a star, likely leaving no remnants. They are strong sources of extra kinetic energy in early galaxies and expected to be detectable as gamma-ray bursts, or perhaps superluminous supernovae in future surveys.

\begin{marginnote}

\entry{GW Chirp} \hfill A gravitational wave signal with rising frequency and amplitude, characteristic of binary object mergers. All signals detected to date have been chirp signals

\entry{GW 190521} \hfill The most massive binary merger observed to date with component masses of 85\,M$_\odot$ and 66\,M$_\odot$ \citep{2020PhRvL.125j1102A}. Their existence challenges stellar evolution models and may imply a dynamical capture pathway.
\end{marginnote}

\subsubsection{Multiple star systems}

The different evolutionary pathways that become available to stars when they are a member of a binary system are more numerous than can be fully described in our short introduction here. We have aimed here to introduce the most important pathways for the massive stars which dominate early galaxies. 

We also note that the evolution of triple stars and higher order multiple systems may also be important. Observations of massive stars in the local Universe indicates that the average number of companions in fact exceeds unity. In general however, most stable triple or higher multiple systems are hierarchical, and the impact on properties of a population will be driven primarily through binary interactions. Truly new and unique pathways that become available in multiple star systems are likely to be rare but important to gaining a full understanding of stellar populations in future \citep[see][]{2021arXiv210804272T}.

\subsection{Supernova and compact remnants}

\subsubsection{Core-collapse and supernovae} 

Supernovae were recognised as dying stars just under a century ago. 
The cataclysmic deaths of stars can inject more energy into the surrounding interstellar medium, via electromagnetic radiation and kinetic energy, than the Sun will emit over its 10 billion year lifetime. Thus while these events only occur once per star they have a disproportionate impact, particularly in clearing their local environs of gas through mechanical feedback.

The majority of supernovae, especially in early galaxies, arise from core-collapse. In massive stars nuclear fusion progresses to a Chandrasekhar mass core of iron-group elements, or oxygen-neon-magnesium. This then collapses to a neutron star or black hole, and releases of the order of $10^{53}$\,ergs of energy, mostly in neutrinos. About 1\% is transferred to the stellar envelope causing it to explode as a core-collapse supernova (CCSN). Dependent on the mass loss that the star has experienced it will produce either a type II (hydrogen-rich) or type Ib/c (hydrogen-free) supernova. In both cases the impact is similar.

There is growing evidence that not every core-collapse necessarily produces a supernova. In some cases, the explosion may be stifled by a newly-forming black hole
and so the star simply stops shining and slowly accretes onto the central black hole \citep{2017MNRAS.468.4968A}. This may reduce the amount of kinetic energy injected into the surrounding environment. Such behaviour may preferentially the case for the stars above  $\sim$20~$M_{\odot}$ and favoured by low metallicities \citep[e.g.][]{2016ApJ...821...38S}. This implies that 
in early galaxies there may be no supernova feedback until after 10~Myrs. 
However all current simulations of this effect have been made assuming the star is a single star. There is some evidence that binary interactions may significantly change the pre-supernova structure of a progenitor and thus make it more explodable \citep{2021arXiv210205036L}, possibly even when a black hole is formed.

Binary interactions may also be responsible for causing more energetic supernovae \citep{2012ARA&A..50..107L}. Binary evolution likely has a role in causing long-gamma-ray bursts and broadline type Ic supernovae \citep[e.g.][]{2020MNRAS.491.3479C}, and evidence suggests they may cause some superluminous supernovae \citep{2021MNRAS.504L..51S}. These  output more than 10 times the typical supernova energy, and tend to arise in low metallicity environments similar to those in early galaxies. 

\begin{marginnote}
\entry{CCSN} \hfill Core-Collapse SN formed at the end of the life of a massive star
\entry{SN Ia} \hfill Thermonuclear SN caused by accretion onto white dwarf
\entry{PISN} \hfill Pair instability supernova occuring in very massive stars
\end{marginnote}

The most extreme types of supernovae are the pair-instability and pulsation-pair instability supernovae. These events occur for stars with helium cores between 64 to 133~M$_{\odot}$ \citep{2002ApJ...567..532H}, which  occur primarily at low metallicities. Here the stellar interior reaches a point where photons are energetic enough to collide and form electron-positron pairs. This reduces the radiation pressure support such that the star collapses and explodes. Above this mass limit the star is completely disrupted, creating many solar masses of nickel-56. Below the limit the star's structure is significantly affected and mass is lost, but it will then evolve towards a normal core collapse. While no such event has been convincingly identified, evidence from the black hole mass distribution of GW transients strongly suggests that these events do happen, with the implications that the early Universe may have been enriched in iron rapidly by such events \citep[e.g][]{2021arXiv210803850I}.

\subsection{Very Massive Stars}\label{sec:VMS}

An important recent advance has already been mentioned a few times above, but we emphasise its importance here: that is an improved appreciation of very massive stars (VMS) above about 100~M$_{\odot}$. While such stars are few in number, their high luminosity means they have a dramatic impact on their surrounding environment through both their high luminosity and their final supernovae, which may be pair-instability events, superluminous supernova, or long-GRBs if they are members of a binary system.

\begin{marginnote}
\entry{Very Massive Star (VMS)}\hfill Star with M$>100$\,M$_\odot$
\end{marginnote}

The existence of such stars is now firmly established \citep{2010MNRAS.408..731C}, although a possibility remains that they might be merger products and hence the result of binary evolution already. Recent studies have found that there is a dearth of young, moderately massive  main-sequence stars \citep{2020A&A...638A.157H,2021A&A...646A.106S}, the reasons for which is still unclear but explanation might be that such stars merge early on in their evolution to create more VMS. 

Wind properties of VMS also remain unclear. They may have very slow winds, unexpected for a hot massive star. Matter in such a slow wind remains close to the star and may be ionized producing strong \heii\ emission lines at low metallicity \citep{2015A&A...578L...2G}.
Finally a peculiarity of the structure of such stars is that they have very large convective cores that take up most of the volume of the star. This means that little mass loss is required before nuclear burning products are exposed at the surface \citep{2020MNRAS.494.3861R} and potentially entrained by stellar winds. This provides another way for these stars to enrich their environments. Further study of this impactful population is clearly needed.

\subsection{Comparing their impacts}

When it comes to comparing the importance of the above factors the most significant factors are rotation and binary interactions. We can see in Figures \ref{fig-hr-single} and \ref{fig-lifetimes} that differences between the different single star tracks considered are relatively minor, although differences in convective core sizes and interior mixing profiles as found by \citet{2021NatAs...5..715P} and \citet{2021arXiv210709075J} may make greater differences. However these variations will be unable to cause variations as large as those due to binary interactions. For example, we can see that interacting binary stars create hot Wolf-Rayet stars at masses that cannot be made by single star winds alone, and the WR lifetime is significantly increased compared to that from any single star models. We note the observational signature of these low mass Wolf-Rayet stars will be different to those typically observed \citep{2018A&A...615A..78G}.

To compare the factors we must consider how they affect the environment of early galaxies. Stars provide energy by ionizing photons, stellar winds and supernovae. The total input is set by the IMF which determines how many stars there are to provide this input. The initial composition or metallicity however does affect the ionizing photon yield and stellar wind input. Lower metallicities imply hotter stars so more ionizing photons but weaker stellar winds. 
Analysing the different models we find that because of these similarities, we can compare which mass range most of the ionizing photons from a stellar population, i.e. weighted by lifetime and IMF, are produced. We find that
stars from 30~M$_{\odot}$ and above produce most of the ionizing photons from a stellar population, while the most massive (100 to 200~M$_{\odot}$) stars can make up to approximately half the number of ionizing photons from a stellar population. This is due to their high luminosities and surface temperatures, despite their short lifetimes.
The distribution of stellar masses that contribute to kinetic energy input from hot star winds extends upwards from the slightly lower initial mass of 10~M$_{\odot}$, but this is highly metallicity dependent. 

In comparison the energy input from core-collapse supernovae is dominated by lower mass stars, in the 8 to $\approx$20~M$_{\odot}$ mass range. This is because, given the approximation that most supernovae release similar total energies, it is the more numerous lower mass stars that dominate. Stars above 20~M$_{\odot}$ only contribute maybe up to 20\% of the energetic feedback from supernovae, unless 
their core-collapse does not have an associated supernovae due to black hole formation. 
From a feedback perspective this is not so important as they contribute the fewest core-collapse events, although we must also note that exotic energetic supernovae types may originate from the most massive stars. These may provide input at younger ages in star formation regions of early galaxies, and in the case of PISN produce large amounts of nickel, they may not have a noticeable impact on galaxy scales compared to the larger CCSN population. 

Thus each feedback mechanism has its own unique mass range that dominates their energy input, varying from the lower mass end dominating supernovae, through a mass range which dominates stellar wind output, up to the highest mass stars dominating ionizing radiation feedback. 
For the most part, these mass ranges do not vary with inclusion of rotation or binary interactions. One detail that does change is the ionizing spectrum: only the hot helium stars that arise from binary interactions can produce significantly harder ionizing spectrum than a population including only single stars.

Finally, if we look further at the events that form black-holes, the mass range that produces most of the black holes is also above 30~M$_{\odot}$, as was the case for the ionizing radiation from a stellar population. Thus studying GW transients and X-ray binaries can give us an insight into those stars that ionized the early Universe.




\begin{summary}[SUMMARY POINTS]
\begin{enumerate}
\item Areas including the initial parameter distribution of stars, the impact of binary evolution and the mass loss rates due to stellar winds have shown significant advances in recent years. Nonetheless significant differences remain between the evolution of stars modelled with different physical assumptions.
\item Observations, ranging from detailed studies of individual stars to the new field of gravitational wave astrophysics, are providing increasingly strong constraints on massive star evolution pathways.
\item Different stellar mass ranges can be traced by different observational indicators.
\end{enumerate}
\end{summary}

\section{MASSIVE STARS IN THE UNIVERSE}

\subsection{Star Formation Histories and Star Formation Rates}
\label{sec:SFRs}

Some of the most fundamental properties of galaxies are their masses and current star formation rates (SFRs), and these are commonly tabulated and released as a `value-added' output of extragalactic surveys. 

Mass can sometimes be determined directly through dynamical properties of galaxies, or through analysis of their lensing potentials. However this requires detailed spectroscopy and is of limited use in large surveys. For most galaxies 
the stellar mass and SFRs are obtained simultaneously from fitting SPS models to the integrated light of the stellar population. Alternatively, scaling relations are calibrated using such fits,
 based on the principle that most of the information regarding a stellar population's age is encoded in its optical colour or ultraviolet flux, while stellar mass most strongly influences a galaxy's bolometric luminosity.

In galaxies with recent star formation, the young high mass stars will dominate the emission, even in the near-infrared. 
In such cases, the instantaneous SFR is often determined from one or more observable star formation rate indicators (SFRIs), such as the strength of the H$\alpha$ emission line. 
For a population forming stars at a constant rate, the line luminosity scales approximately linearly with SFR and the line equivalent width scales with stellar population age \citep[see][]{2012ARA&A..50..531K}. Similar relations have been established for the ultraviolet continuum, the far-infrared continuum,
and the radio synchrotron continuum.
More recently, the X-ray continuum emission from accreting compact remnants in galaxies has also been identified as an SFRI. In each case, the correlation with SFR arises from a combination of the emmissive luminosity and the lifetimes of massive stars. From these, the complementary population of lower mass stars is determined from an SPS model. 

\begin{figure}
    \centering
    \begin{tabular}{p{0.48\textwidth}p{0.48\textwidth}}
    \includegraphics[width=0.45\textwidth]{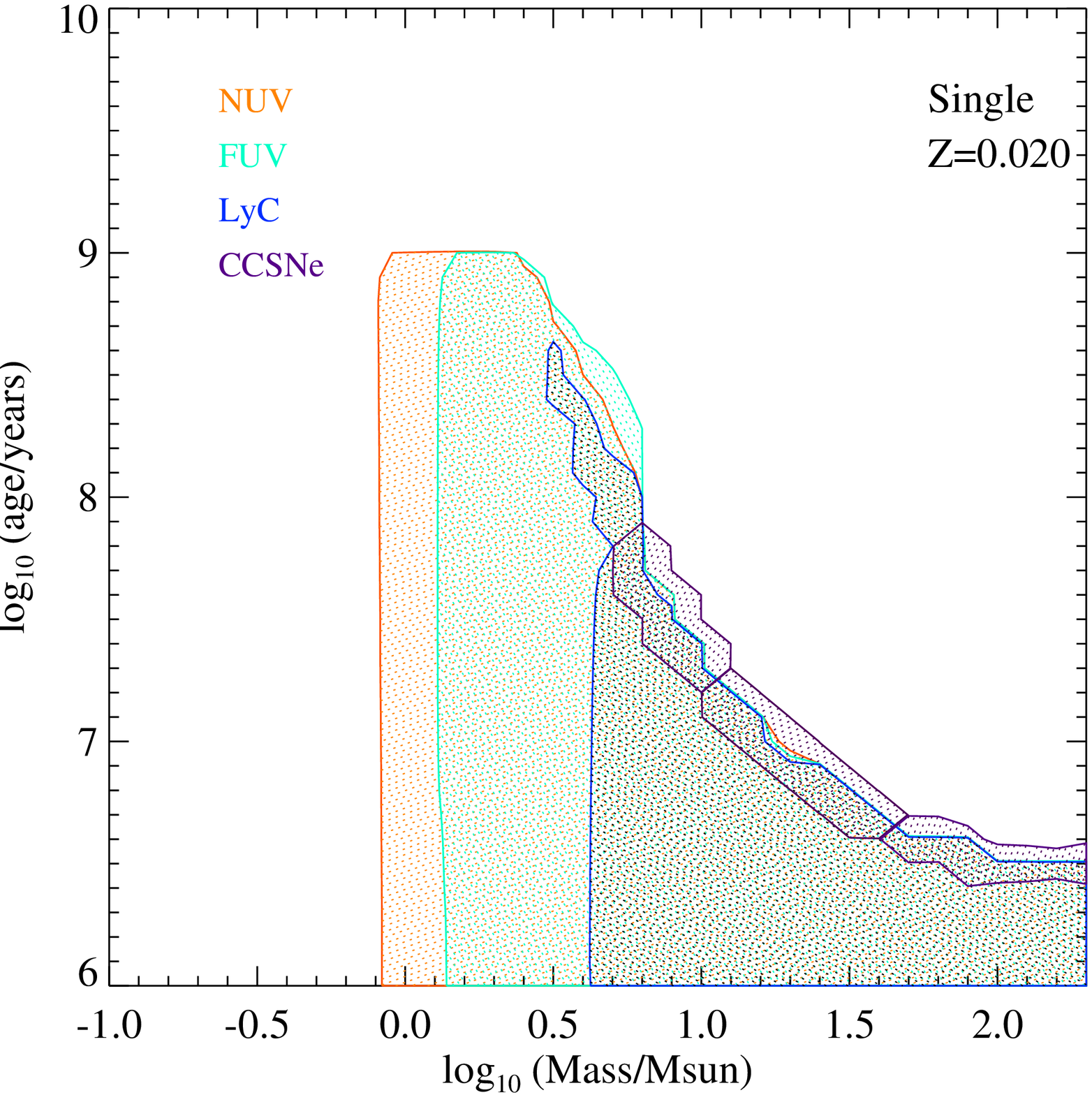} &
    \includegraphics[width=0.45\textwidth]{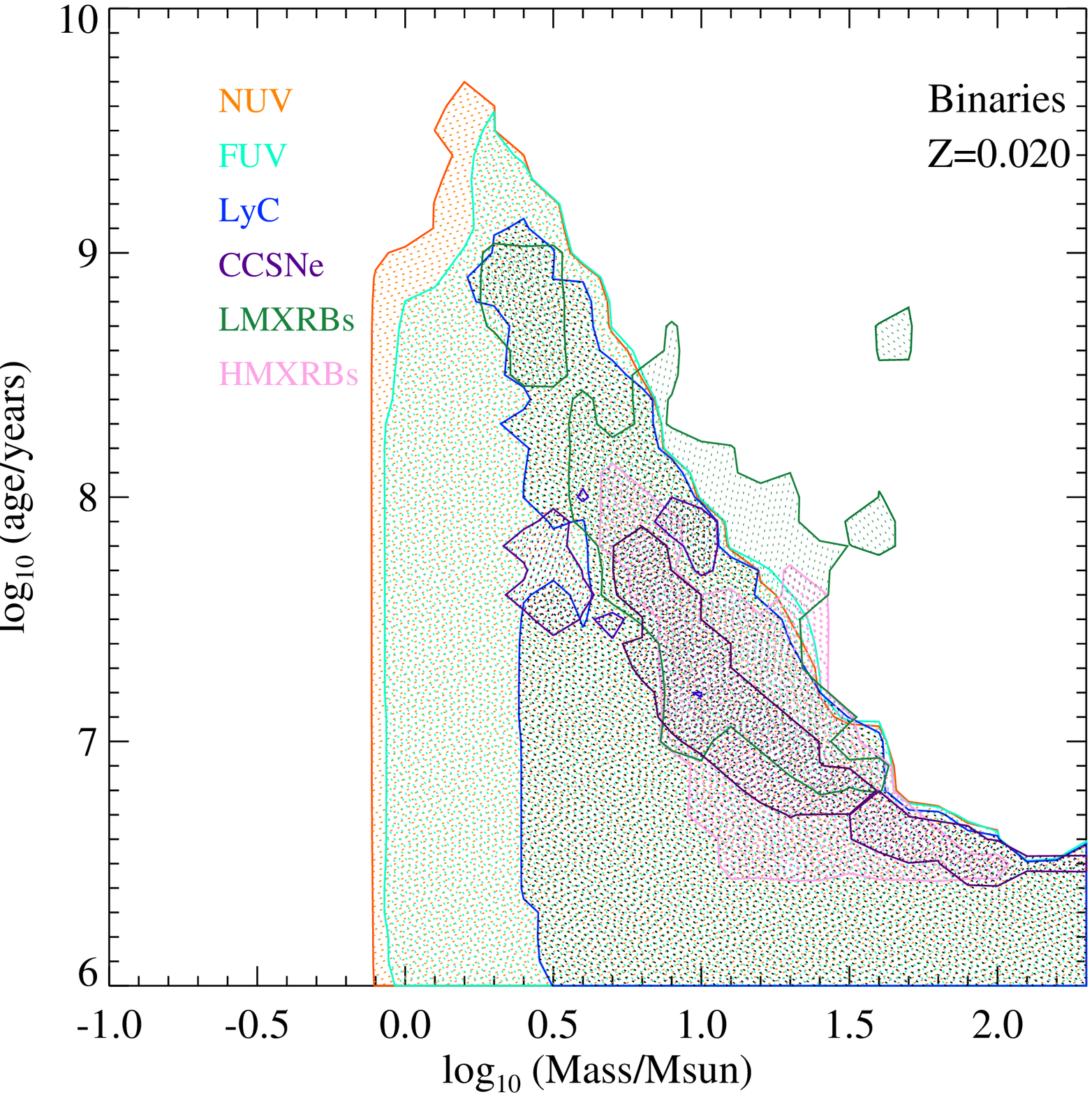} \\
    \includegraphics[width=0.45\textwidth]{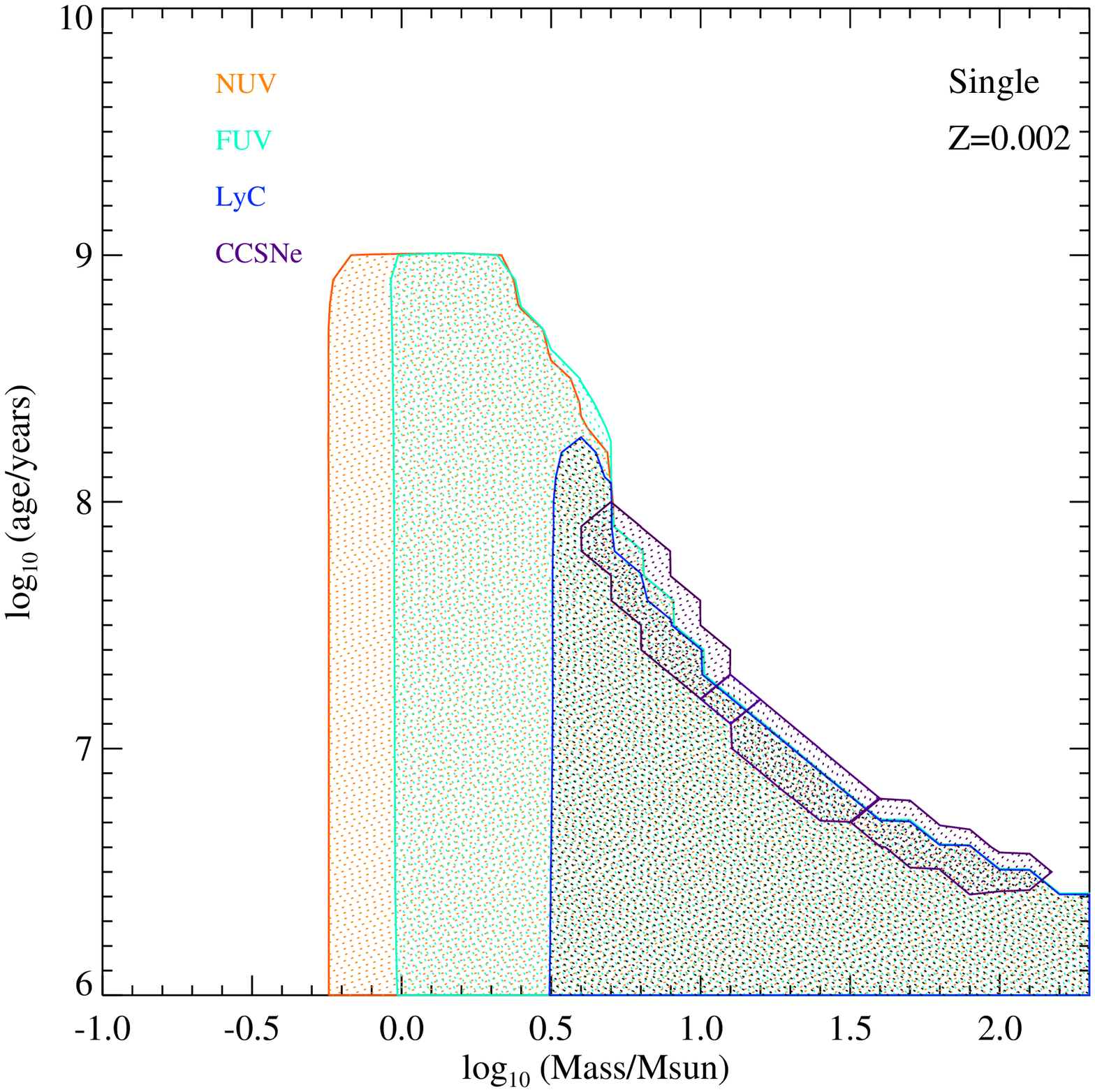} &
   \includegraphics[width=0.45\textwidth]{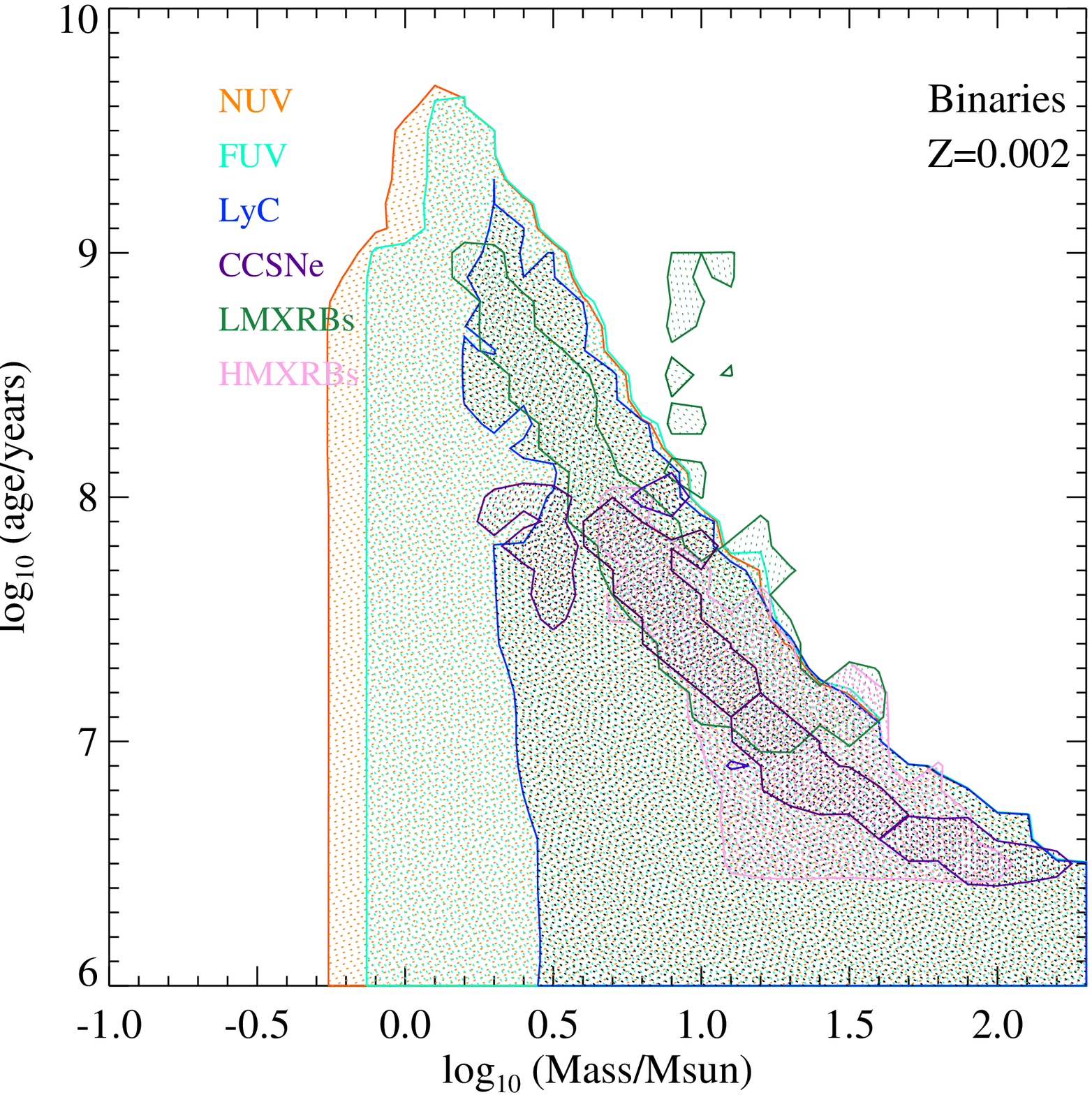}
    \end{tabular}
    \caption{The masses and timescales of stars contributing to key star formation rate indicators in the absence (left) or presence (right) of binary stellar evolution pathways. Top: models at Z=0.020, bottom: Z=0.002. 
    }
    \label{fig:SFRIs}
\end{figure}

To demonstrate the impact of key massive stellar population assumptions, Figure \ref{fig:SFRIs} shows the time scales and mass ranges of stars contributing to specific SFRIs in the presence or absence of binary stellar evolution pathways, for models at $Z=0.002$ and 0.020. 
Mass ranges and ages for which the stellar population in a $\Delta(\log(M/M_\odot))=0.1$, $\Delta(\log(\mathrm{age/years}))=0.1$ interval exceeds certain thresholds are shaded, with the thresholds held constant between binary and single star populations. Assuming our fiducial initial stellar population with a total mass of $10^6$\,M$_\odot$, 
we set a threshold of $10^{46}$ s$^{-1}$ for contribution to ionizing photon production, a continuum luminosity $L=10^{32}$\,ergs\,s$^{-1}$\,\AA\ in the far ultraviolet (FUV) and near ultraviolet (NUV) photometric bands, and require 10 core collapse supernovae (CCSNe). For binary populations we also calculate the gravitational energy released by accretion onto high mass and low mass X-ray binaries (HMXRBs and LMXRBs), setting a threshold for time-averaged energy release $10^2$\,L$_\odot$. These thresholds are arbitrarily selected for illustrative purposes. 

\begin{marginnote}
\entry{Commonly used SFRIs} \ 

\entry{UV continuum} \hfill traces hot but non-ionizing stellar photospheres.
\entry{H$\alpha$ emission line} \hfill traces photons with $h \nu>$13.6\,eV.
\entry{FIR continuum} \hfill traces dust heated by absorption of UV radiation.
\entry{Radio continuum} \hfill traces emission from electrons accelerated by supernovae.
\entry{X-Ray luminosity} \hfill traces presence of accreting compact objects.

\bigskip

See \citet{2012ARA&A..50..531K} for more details.
\end{marginnote}

The ultraviolet continuum luminosity is used primarily in the context of spectral energy distribution fitting and varies  linearly with the SFR in a stellar population \citep[see][]{2012ARA&A..50..531K}, as long as constant star formation extends over a timescale equal to or greater than the lifetime of the lowest mass contributing star. As Figure \ref{fig:SFRIs} demonstrates, the (mass dependant) binary fraction in a population can  complicate this interpretation. The ultraviolet-luminous lifetime of stars is extended at all masses, and the linear calibration between SFR and ultraviolet luminosity breaks down unless constant star formation is maintained for substantially longer intervals than hitherto assumed. 

The same trends are seen in ionizing photon production, which is 
usually inferred based on either the ultraviolet continuum or through spectroscopy of the nebular emission lines powered by the ionizing flux. Again, ionizing flux takes longer to stabilise, affecting the calibration of H$\alpha$ luminosity as an SFRI, and the line equivalent width as an age indicator \citep[e.g.][]{2018A&A...613A..35K}.

An indirect measurement of the SFR is obtained from the synchrotron continuum at radio wavelengths. This emission arises from highly energetic electrons accelerated by magnetic fields associated with supernovae and their remnants \citep{1992ARA&A..30..575C} and so scales with the core-collapse supernova rate.
The use of radio emission as an SFRI is calibrated by empirical relations designed to ensure consistency with other indicators. 

X-ray luminosity has also been empirically calibrated for use as an SFRI. 
The faint end of the X-ray luminosity function is dominated by star forming galaxies 
in which mass accretes onto a compact object due to Roche-lobe overflow from a companion (i.e. XRBs). 
As Figure \ref{fig:SFRIs} illustrates, the accretion luminosity of a large population traces relatively massive stars and young stellar populations. A complication arises however from the stochastic nature of accretion. 
While we calculate time-averaged values, the emission at any one epoch is 
dominated by a handful of highly luminous objects undergoing short lived episodes of accretion, while many other binaries are quiescent. Thus this is a statistical method for estimating SFRs, rather than one which can be used to interpret the flux of any one system or the numbers of individual X-ray luminous sources.

However the results shown in Figure \ref{fig:SFRIs} are for just one stellar population with specified initial mass function, binary parameters and selected metallicity, and constructed using models with a limited consideration of rotation. Inevitably, changing any of these parameters also alters the temperatures, lifetimes and interaction probabilities of stars, modifying the relative calibration of SFRIs and parameters derived from them. 
Substantial uncertainty in the calibration of masses and SFRs (whether using fixed relations or SED fitting) also arises from variability in a galaxy's star formation history.  Many galaxies, particularly in the distant Universe, exhibit bursty star formation which cannot be captured by a parametric star formation history  \citep[e.g.][]{2015MNRAS.452.1447K,2019ApJ...876....3L}. 
A galaxy with bursty star formation may show emission line spectra consistent with zero age, or a high apparent mass unless the star formation history is taken fully into account.

\subsection{The Distant Universe as a testbed}

An area in which uncertainties regarding the impact of massive stars has come to the fore in recent years is in the study of the distant Universe. 
Stellar populations at high redshift are often first identified by their ultraviolet emission, modified by absorption due to neutral hydrogen in the intergalactic medium to show a distinctive Lyman break \citep{1990ApJ...357L...9G, 1995AJ....110.2519S}. 
This, combined with the short timescales available for galaxy evolution at high redshift, means that the stellar populations observed in the distant Universe are typically very young, and their emission is dominated by massive stars. Other recent reviews provide a thorough survey of the high redshift universe \citep[e.g.][]{2020ARA&A..58..617O,2020ARA&A..58..661F}. 
Here we focus on properties which are either affected by, or affect in turn, our understanding of the evolution of massive stars.

Cosmic noon is a name given to the epoch at which
the comoving space density of ultraviolet photons was higher than at any time since the Big Bang \citep[see][]{2014ARA&A..52..415M}. 
By contrast, cosmic dawn marks the emergence of the first stars, illuminating the intergalactic medium (IGM). Both the start and the end of this period are rather ill defined. The first, metal-free stars (known as Population III) were probably born around $z\sim25$, with metal-enriched Population II stars following within a few million years, as the Population III supernovae pollute their surroundings. The end of the epoch coincides with the end of the epoch of reionization around $z\sim5-6$. 
The stellar populations at these times formed when the Universe was yet to reach its current level of metal enrichment. Processes with long delay times, such as iron formation in type Ia supernovae, cannot have operated, and the composition of nebula gas will be dominated by the products of rapid processes. 
As a result, galaxies in the distant Universe were not only lower in overall metallicity but also potentially very different in elemental abundant ratios to those observed nearby.

These factors combine to make the distant Universe a natural testbed for our understanding of massive star evolution. Any uncertainties in the effects of stellar initial conditions or evolution, whether through stellar spectra, temperatures, abundance influences on the IMF or binary population parameters, will translate into uncertainties in interpretation and analysis of electromagnetic emission from distant galaxies. In theory then, stellar populations observed in the distant Universe could place tight constraints on the validity of stellar population models.

\begin{marginnote}
\entry{Cosmic Noon} \hfill The epoch of peak cosmic star formation and AGN accretion. Variously defined in the literature as $z\sim1-2$, $2<z<3$, $1.5<z<3$ or $z\sim2$.
\entry{Cosmic Dawn} \hfill The epoch at which the first luminous objects began to shine, typically applied to $z>6$
\entry{Cosmic Brunch} \hfill Slightly tongue-in-cheek term used to refer to the intervening epoch at around $3<z<5$.
\end{marginnote}


\subsection{The Epoch of Reionization}
\label{sec:reion}

The intergalactic medium is formed from a complex cosmic web of gas clouds. These are largely comprised of hydrogen gas, whether unprocessed since the big bang nucleosynthesis, or enriched to some extent by metals expelled in winds from galaxies. In the immediate aftermath of recombination (when protons and electrons first combined to form atoms), the intergalactic medium was predominantly neutral. 
At the current time, however, the intergalactic medium is predominantly ionized, with only the densest clouds remaining predominantly neutral. The phase transition between these neutral and ionized states occurred during the epoch of reionization (EoR). 

Observational evidence has now provided reasonably tight constraints on the progression of reionization as traced by the volume-averaged fraction of neutral hydrogen in the ISM. 
A combination of all available observational signatures has now produced a clear picture of a patchy, gradual reionization which progressed through ionized bubble overlap, beginning at around $z\sim10-12$ \citep{2020A&A...641A...6P} and only reaching completion around $z\sim5-5.5$ \citep[e.g.][]{2020MNRAS.491.1736K,2021arXiv210803699B}. 
In the intervening time some quasars and gamma ray bursts show evidence for substantial ultraviolet photon absorption by neutral gas while others do not. However a true measurement of the patchiness of ionized gas in the Universe and its evolution over time 
awaits constraints on the power spectrum of the 21cm radiation in the EoR 
from instrumentation such as the Square Kilometer Array.

Such a power spectrum will clarify the source of photons for reionization \citep[e.g.][]{2016ARA&A..54..313M}. 
However the balance of evidence at the current time favours ionization by abundant, relatively faint sources: star forming galaxies. The estimated number density of such galaxies  now appears to be sufficiently high to provide the ionizing radiation required - given certain assumptions \citep[see e.g.][ and references therein]{2020MNRAS.498.2001M}. 
Amongst those, two key parameters are the strength of the ionizing radiation field associated with galaxies, and the fraction of ionizing photons produced which are able to escape their galaxy. 
These can (at present) only be based on our understanding of massive stars.

\subsection{The Ionizing Photon Production Efficiency, \xiion}\label{sec:xiion}

The ionizing photons responsible for reionizing the universe are those emitted at rest-frame wavelengths shortwards of 912\AA. Due to the long sightlines to the distant Universe, these wavelengths are totally absorbed by intervening neutral hydrogen gas. 
As a result, the ionizing continuum can only be observed directly in very rare cases known as Lyman continuum leakers. Instead, observations taken longwards of the Lyman break are typically used to infer the strength of ionizing radiation. 

A direct constraint on the ionizing radiation field can be obtained from rest-frame optical recombination lines, including the Balmer series of hydrogen. The ratio of the Balmer lines to the ionizing photon density is only weakly dependent on nebular gas conditions or ionizing emission spectrum \citep{2006agna.book.....O}, allowing a direct conversion of $L(H\alpha)$ to $Q(H0)$ (see Supplementary Information). 
However such observations are challenging at high redshifts, where the Balmer lines are shifted into the observed-infrared. Instead, estimates can be based on the Lyman-$\alpha$ line, requiring the assumption of a further relationship $L(Ly\alpha)/L(H\alpha)$  with a canonical value of 8.7 \citep{2006agna.book.....O}. However this is less than satisfactory: the Lyman-$\alpha$ line is resonantly scattered and far more susceptible than the Balmer line to  extinction by dust or nebular gas. 
Such estimates also cannot account for uncertainties such as the gas covering fraction, or fraction of the emitted photons escaping into the intergalactic medium rather than being reprocessed by nebular gas (see next section). 

An alternate approach, widely used at high redshift, is to use observations taken in the far-ultraviolet continuum as a proxy. The ionizing photon density is inferred from the continuum luminosity density at 1500\,\AA\ using SPS models to fix an ionizing photon production efficiency, $\xi_\mathrm{ion,0}$:

\begin{equation} 
\xi_\mathrm{ion} [\mathrm{ergs}^{-1}\,\mathrm{Hz}]= (1-f_\mathrm{esc})\ \xi_\mathrm{ion,0} = \frac{(1-f_\mathrm{esc})\ Q(H0) \,[\mathrm{s}^{-1}] }{ L_{1500}\, [\mathrm{ergs}\,\mathrm{s}^{-1}\,\mathrm{Hz}^{-1}]}.
\end{equation}

Here, $f_\mathrm{esc}$ is the fraction of ionizing photons which escape from the galaxy into the intergalactic medium, and $L_{1500}$ is the continuum luminosity at 1500\AA\ in the rest frame. 
\xiion\ and \xiion$_{,0}$ have been used rather inconsistently in the literature, with many quoted values for \xiion\ implicitly assuming an escape fraction $f_\mathrm{esc}=0$ (i.e. that all the ionizing photons produced are reprocessed by nebular gas). 

\begin{marginnote}
\entry{The escape of ionizing photons} \hfill Most photons energetic enough to ionize hydrogen are absorbed in a galaxy's interstellar medium. The fraction that can escape the galaxy without being absorbed and reprocessed to lower energy is known as {$f_\mathrm{esc}$}.

\entry{Ionizing Photon Production Efficiency} \hfill relates the strength of the ionizing continuum (which can't be observed directly) to the observable continuum luminosity in the far-ultraviolet. Strongly dependant on the hottest stars in a population.

\entry{$\mathbf{Q(H0)}$} \hfill The total production rate of photons with $h \nu > 13.6$\,eV from an integrated stellar population.
\end{marginnote}

 This parameterisation allows number densities and luminosity functions of observed Lyman break galaxies to be converted first to luminosity densities in the UV continuum and then to inferred ionizing photon densities, which can be compared to those required to reionize the Universe. Given reasonable assumptions, star forming galaxies are capable of reionizing the Universe if \xiion\ is typically high. For example, \citep{2021arXiv210510518D} estimate that galaxies can reionize the Universe if $\log_{10}f_{\rm esc}\xi_{\rm ion}/\mathrm{(erg/Hz)}^{-1} =25.02_{-0.21}^{+0.45}$ at $z\sim6$, and other comparable estimates exist in the literature in which either \xiion\ or $f_\mathrm{esc}$ has been fixed rather than calculating a joint constraint \citep[e.g.][]{2019ApJ...879...36F}. 

\begin{figure}
    \centering
    \begin{tabular}{p{0.9\textwidth}}
   \includegraphics[height=0.57\textwidth]{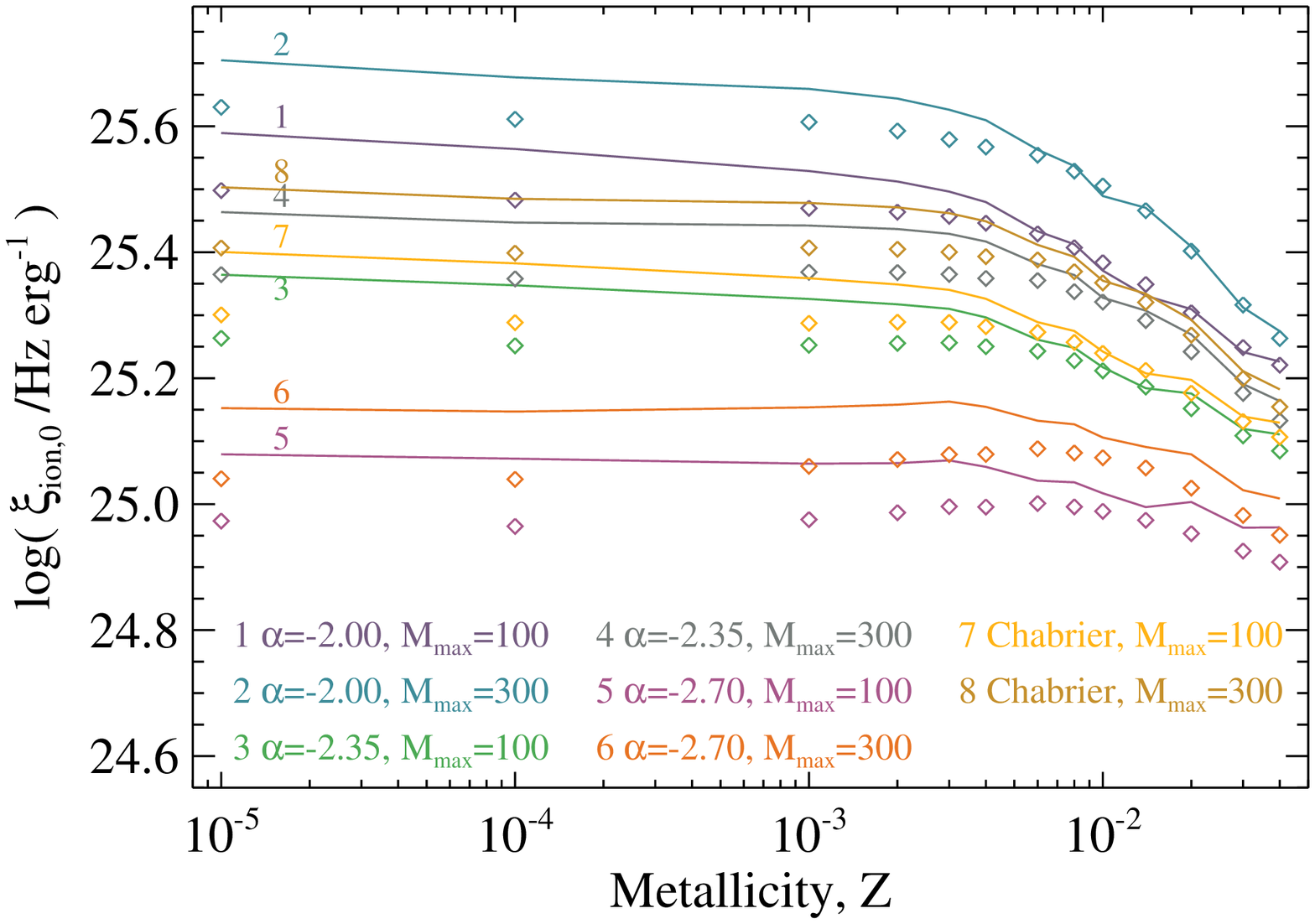} \\
    \includegraphics[height=0.57\textwidth]{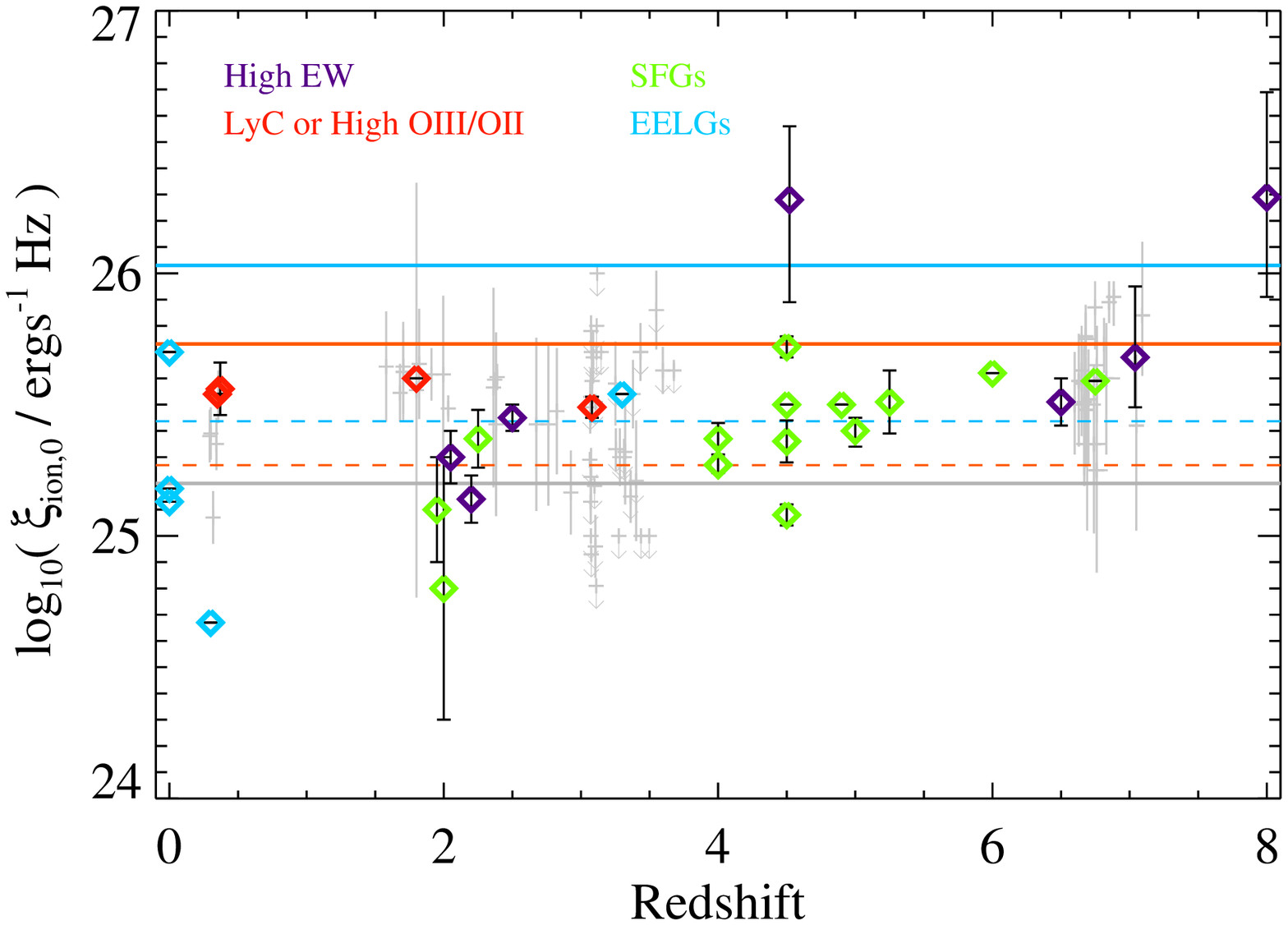} 
    \end{tabular}
    \caption{The theoretical ionizing photon production efficiency, $\xi_{\mathrm{ion},0}$, for model stellar populations as a function of metallicity and assumed initial mass function. 
    Top: For constant star formation, binary population models are shown with solid lines and single star populations with diamond symbols.  IMFs 1-6 are broken power laws with a slope of -1.3 from 0.1 to 0.5\,M$_\odot$ and a slope $-\alpha$ between 0.5\,M$_\odot$ and M$_\mathrm{max}$. IMFs 7 and 8 have $\alpha=2.35$, but the low mass break in the power law is replaced by the low mass roll-over prescription of \citet{2003PASP..115..763C}.
    Bottom: A compilation of observational estimates of \xiion$_{,0}$ drawn from the literature (see Section \ref{sec:xiion} for references). Coloured points show estimates based on samples of two or more objects. 
    Grey points show selected samples of individual galaxies. 
    Horizontal lines indicate the maximum values at Z=0.020 (orange) and Z=0.002 (blue), for a zero age starburst (solid) and constant star formation (dashed), and also the value suggested by \citet[][grey]{2013ApJ...768...71R}. 
    }
    \label{fig:xiion_model}\label{fig:xiion_data}
\end{figure}

Figure \ref{fig:xiion_model} (left) shows the expected theoretical $\xi_\mathrm{ion,0}$ for populations which form stars at a constant rate for an interval of $10^9$\,years. 
At Solar and super-Solar metallicities the effect of IMF and binary pathways is relatively small, with all estimates of \xiion\ consistent within 0.2\,dex. However at very low metallicities (Z=$10^{-5}$), the same variation in population parameters leads to a divergence 
of 0.8\,dex, with the upper mass limit and the inclusion or omission of binaries dominating the uncertainties. 
Binary evolution pathways boost the ionizing photon efficiency by $\sim0.1$\,dex at low metallicities, while having a negligible effect for near-Solar populations. This boost occurs not in the peak value of ionizing photon efficiency at near-zero age, but in the duration of this peak and hence the integrated light of mixed-age populations \citep[see also][]{2019ApJ...882..182C}. We stress that 
this span does not necessarily account for the true uncertainty on \xiion\ since we restrict ourselves to population statistics uncertainties and do not vary most of the uncertainties in stellar model physics (e.g. winds or rotation) discussed in section \ref{sec:stellar}.

Since the advent of highly-multiplexed multi-object spectroscopy in the distant Universe, a growing sample of galaxies exists for which both the rest-optical 
emission lines and the rest-ultraviolet continuum luminosity density can be measured. 
These can be compared to obtain a direct constraint on $\xi_\mathrm{ion}$ (assuming that the same stellar population is being measured and that escaping photons and absorption due to dust have been correctly accounted for). 

In Figure \ref{fig:xiion_data} (right) we present a compilation of $\xi_\mathrm{ion}$ estimates drawn from the literature. Wherever possible we have shown average measurements based on large samples of objects, and we quote the \xiion$_{,0}$ as calibrated by the original authors. Where no data for \fesc\ was provided, we assume \fesc$=0.$ (a number of authors themselves assume this). Measurements of this quantity have, perhaps naturally, been concentrated on the most highly ionizing and highly luminous objects at a given redshift - indeed those taken in the very distant Universe are only possible for extremely powerful emission line galaxies. 
They have also been obtained for samples obtained through a variety of selection methods, which produce overlapping samples. For clarity, in Figure \ref{fig:xiion_data} we attempt to classify the samples into four broad categories. In purple, we show reported sample means for galaxies selected as very luminous examples of narrowband or spectroscopically-selected emission line galaxies \citep[typically H$\alpha$ emitters at cosmic noon and Ly$\alpha$ emitters at high redshift,][]{2020MNRAS.493.5120M,2017MNRAS.464..469S,2019MNRAS.489.2355D,2018A&A...619A.136M,2021MNRAS.505.1382M}. 
Cyan symbols indicate galaxies selected as extreme emission line galaxies, either based on broadband photometry excesses or directly through spectroscopy \citep[EELGs, ][]{2020ApJ...904..180O,2019ApJ...882..182C,2019ApJ...878L...3B,2016A&A...591L...8S,2018ApJ...859..164B}. Red symbols show samples of galaxies identified as Lyman Continuum leakers (i.e. with a measured high \fesc) or as having very high [OIII]/[OII] ratios \citep{2018A&A...616L..14S,2019MNRAS.489.2572T,2021MNRAS.503.4105T,2020MNRAS.497.4293G}. Other star forming galaxy samples, including both Lyman break galaxies and typical Lyman alpha emitters are shown in green \citep{2016ApJ...831..176B,2011ApJ...738...69S,2016MNRAS.460.3587M,2020ApJ...895..116E,2020ApJ...889..180N,2021MNRAS.502.6044E,2019ApJ...884..133F,2019A&A...627A.164L,2018ApJ...859...84H,2018ApJ...855...42S,2020ApJ...892..109N}. To illustrate the scatter within samples, we also shown individual galaxy measurements for an EELG sample   \citep{2019ApJ...882..182C} at $z\sim1.6-3$,  Lyman continuum leakers \citep{2018A&A...616L..14S} at $z\sim0.3$,  high [OIII/OII] galaxies \citep{2016ApJ...831L...9N} at $z\sim3$ and emission line galaxy samples   at $z>6.5$ \citep{2021MNRAS.502.6044E} and  $z\sim3.1-3.6$ \citep{2020ApJ...904..180O}. These are shown as small, grey symbols.

The compiled sample data shows little evidence for evolution with redshift, although a number of authors \citep[e.g.][]{2020ApJ...889..180N} have noted a  dependence within any given sample on stellar mass and luminosity, and on the ultraviolet continuum spectral slope. These correlations are all likely determined by underlying dependence of \xiion\ on age and metallicity.  Blue spectral slopes are associated with both young stellar ages and low metallicity populations. High stellar mass or luminosity is typically correlated with more mature stellar populations and higher metallicities. 

Horizontal lines on Figure \ref{fig:xiion_data} indicate selected theoretical values for \xiion$_{,0}$, specifically zero age (solid) and constant star formation case (dashed) ionizing photon efficiencies predicted in our fiducial binary models at Z=0.020 (orange) and Z=0.002 (blue), and (in grey) the widely-used log(\xiion)=25.2 value derived by by \citet{2013ApJ...768...71R}, based on constant star formation SPS models of \citet{2003MNRAS.344.1000B}. 
The majority of measurements of extreme objects, both at high and low redshifts, exceed both the \citet{2013ApJ...768...71R} prediction and the similar Solar metallicity, constant star formation case using our fiducial model. The majority of large samples of star forming galaxies, which include Lyman break galaxies with weak line emission,  yield values consistent with constant star formation at lower metallicity. However a sizeable fraction of measurements, particularly at the highest redshifts, lie well above the value expected for constant star formation, and approach that for a zero age starburst, consistent with bursty or rising star formation histories.

\begin{marginnote}
Measured values of \xiion\ frequently exceed \xiion$=25.2$ suggesting the presence of a hard ionizing radiation field.
\end{marginnote}

Our understanding of \xiion\ in the distant Universe still remains at an early stage, due to the very limited samples of galaxies with sufficiently strong line emission for robust, model-independent measurements to be made, the still more limited sample with tight constraints on both star formation history and metallicity, and the necessity of treating spatially unresolved integrated spectra rather than individual star forming regions. On the theoretical side, the far-ultraviolet spectra of binary evolution products and low metallicity, very massive stars is an area which certainly requires further work, with significant differences between the different stellar atmosphere models \citep[e.g.][]{2018A&A...615A..78G} which inform population synthesis. The role of rapid rotation in stellar evolution may also be significant \citep{2015ApJ...800...97T,2017PASA...34...58E}, particularly at the very low metallicities which may be typical at cosmic dawn. However current estimates certainly suggest the efficient production of ionizing photons which is expected at sub-Solar metallicities, and which may be explained as a result of the high inferred binary fraction amongst massive stars \citep{2014MNRAS.444.3466S, 2019A&A...629A.134G}.

\subsection{The escape fraction of Lyman continuum photons, \fesc}

An alternative to increasing the inferred photon production efficiency is to permit a higher fraction of the ionizing photons that are produced to escape from the environs of stellar populations into the intergalactic medium without extensive reprocessing and thermalisation of their energy by gas or dust.  The escape fraction can be defined in a variety of functionally-similar but technically different ways. Theoretically the escape fraction is given by the ratio of ionizing photons escaping a galaxy's potential well to ionizing photons generated by a stellar population.
In the context of the distant Universe, a relative escape fraction, \fesc = \fesc$_{,LyC}$/ \fesc$_{,UV}$ is commonly used, where \fesc$_{,UV}$ is the fraction of emitted photons escaping their host galaxy in the ultraviolet continuum and \fesc$_{,LyC}$ is the same quantity for ionizing photons \citep{2001ApJ...546..665S}. This scaling permits an estimate of the ionizing flux which is somewhat less dependent on the dust extinction in high redshift galaxies. 

The escape fraction in galaxies is likely a complex function of time, geometry and gas density. It is not directly determined by their stellar populations, and so cannot be predicted by a population synthesis model, but simulations demonstrate that it is nonetheless strongly affected by the massive star population which modifies the gas conditions in a given galaxy. 
Constraining the escape fraction of Lyman continuum photons observationally is extremely challenging.  Deep observations of either spectra or photometry are obtained below the source Lyman limit (912\AA\ rest). These are then compared to the expected ionizing photon production - again based on SPS modelling - to infer the fraction of expected photons which are actually observed. The results usually have a strong dependence on the assumed \xiion$_{,0}$ value for the population, although in some cases these parameters are fitted simultaneously.  A low assumed \xiion$_{,0}$ will lead to a high \fesc\ estimate if flux is detected shortwards of 921\AA.

In rare examples, galaxies known as Lyman continuum leakers, \fesc\ may be sufficiently high for the continuum luminosity shortward of 912\AA (rest) to be measured directly. The first local star forming galaxy identified as a Lyman continuum leaker was Haro 11, a 0.2\,Z$_\odot$ blue compact dwarf which had already been identified as a potential analogue for galaxies in the distant Universe. The leakage in this source was not extreme, with $0.04 < $\fesc$ < 0.10$ \citep{2006A&A...448..513B}, and other similar sources also showed \fesc$<0.1$ \citep[e.g.][]{2013A&A...553A.106L,2016ApJ...823...64L}. 
At around the same time, the first individual galaxies (2 from a sample of 14 Lyman break-selected sources) were identified which showed Lyman continuum leakage at $z\sim3$, with \fesc$=0.65$ and \fesc$\sim1$, compared to \fesc$\sim0.11-0.18$ for a stack of the complete sample \citep{2006ApJ...651..688S}. 

While some early claims have been modified in the light of deeper data, there is now a modest but growing number of galaxies for which individual escape fraction detections \fesc$>0.5$ have been identified. The most extreme examples have mostly been identified around $z\sim3$ where the effect of redshift is to place the Lyman continuum into a sensitive optical band. These include Ion2 \citep[z=3.2,][]{2015A&A...576A.116V}, Q1549-C25 \citep[z=3.2,][]{2016ApJ...826L..24S}, Ion3 \citep[z=4,][]{2018MNRAS.476L..15V}, and the gold sample of \citet{2019ApJ...878...87F} at $z=3.1-3.6$. In the more local Universe, some tens of objects have now been identified with \fesc$\sim0.1-0.3$ \citep{2021MNRAS.503.1734I,2020MNRAS.497.4293G,2021ApJ...916....3W}. There have also been a number of proxies for \fesc\ proposed including the Lyman-$\alpha$ emission line profile \citep{2015A&A...578A...7V} and the strength of low ionization absorption lines in the ultraviolet \citep{2015ApJ...810..104A,2018A&A...616A..30C,2021ApJ...916....3W}. 

Escape fraction measurements are frequently carried out on stacks of galaxies, in order to improve the signal-to-noise of a Lyman continuum detection, and permit the average galaxy properties to be estimated. While early estimates of the escape fraction from high redshift galaxies were relatively high \citep{2001ApJ...546..665S}, the majority of recent studies have suggested that 
the average \fesc\ of star forming galaxies seldom exceeds a few percent, either in the local Universe \citep[e.g][]{2009ApJS..181..272G}, around cosmic noon \citep[e.g][]{2016ApJ...819...81R, 2020ApJ...904...59A} or at high redshifts \citep[e.g][]{2017A&A...601A..73M,2021MNRAS.505.2447P}. 
A key recent study is that of \citet{2018ApJ...869..123S}, which includes a number of individual spectroscopic detections of leaking Lyman continuum photons at $z\sim3$, as well as analysis of stacked spectroscopic data for a sample of galaxies at the same redshift. They find \fesc$=0.09\pm0.01$ after taking into account the stochastic effect of absorption in the foreground IGM, which complicates such measurements by inducing a strong variation in continuum opacity unassociated with the galaxy itself. This results in a bias since observations of only galaxies with detections select preferential paths through the IGM with low continuum opacity such that it is difficult to correct for the sample mean absorption due to the foreground IGM. 

\begin{marginnote}
Measured $f_\mathrm{esc}$ values in galaxy samples are typically $<10$\% although some individual examples show up to 100\% escape.
\end{marginnote}

Given that sources with low extinction are more ultraviolet luminous and so more easily studied, one might expect measured estimates of escape fractions to be biased high compared to the globally averaged population. 
Studies of radiative transfer through the interstellar medium of galaxies \citep[e.g.][]{2008ApJ...672..765G,2006ApJ...651L..89R} have shown that the effect of even small column densities of gas or dust on the ionizing spectrum is significant, and if the interstellar medium is distributed fairly uniformly, the expected escape fractions is just a few percent along most sightlines \citep[e.g.][]{2009ApJ...704.1640L}. As a result, unless low density sightlines can be found, insufficient photons escape galaxies to reionize the Universe. 

Hence simulation and theoretical efforts have focussed on the possibility that channels can be blown almost entirely clear of gas by injections of kinetic energy \citep[e.g.][]{2004ApJ...613L..97M,2009MNRAS.398..715Y} 
either from the strong winds of very young stellar populations or from the shockwaves of supernovae as the most massive stars end their lives \citep{2001ApJ...558...56H}.
These channels then provide paths through the circumstellar and circumgalactic medium and into the IGM with an \fesc\ of unity, permitting Lyman continuum photons to escape and reionize the Universe. This picture is supported by observational evidence from the distant Universe which suggests a close correlation between the covering fraction of neutral gas in a galaxy and its ionizing radiation properties \citep[e.g.][]{2021arXiv210805363R}.

Crucially, such a paradigm requires channels to be cleared through the gas while hot young stars are still active to power a strong Lyman continuum. 
By the time sufficient kinetic energy is injected to clear the ISM, a single star population has likely exhausted the majority of its ionizing photon production. This delay constraint can be seen in the left hand panels of Figure \ref{fig:SFRIs}: core collapse supernovae occur in a narrow time window after the onset of star formation, with very few ionizing photons emitted substantially afterwards  As a result while the escape fraction at late times may be high, the number of escaping photons remains low and galaxies may not be able to reionize their surroundings until times significantly later than required by observational constraints \citep{2016MNRAS.459.3614M}. 

Improvements in modelling massive stars, and in particular the introduction of binary evolution pathways and very massive stars, have made a significant impact in the modelling of reionization \citep{2016MNRAS.459.3614M,2020MNRAS.498.2001M,2018MNRAS.479..994R,2021MNRAS.505.2207D}. Binary populations continue to generate an ionizing photon continuum well after the first, highly-energetic, supernovae occur, as the right hand panels of Figure \ref{fig:SFRIs} demonstrate, through rejuvenated and stripped binary components. As a result, binary populations may be able to clear their own channels for ionizing photon escape \citep[although c.f.][]{2021MNRAS.503.3698H}. 

\begin{marginnote}
Mechanisms which prolong ionizing photon production, such as binary interactions or perhaps rotation, may be key to allowing supernovae to open Lyman continuum escape channels.
\end{marginnote}

While most of the reionization simulations to incorporate the effects of binary evolution to date have used BPASS stellar population models to inform their ionizing photon production rate and feedback prescription, we note with interest the recent work of \citet{2020ApJ...901...72S} which used the independent binary ionization models of \citet{2018A&A...615A..78G,2019A&A...629A.134G} and reached a similar conclusion.

\subsection{The Ionizing Spectra of Galaxies}\label{sec:spec}

Estimates of the ionizing photon efficiency and Lyman continuum escape fraction are most sensitive to photons emitted at wavelengths just short of the hydrogen ionization edge at 912\AA. However 
the shape of the ionizing spectrum emitted by galaxies 
is most sensitively probed by an indirect method: analysis of the nebular line emission spectrum of galaxies. 
The relative strength of different lines provides a constraint on the abundance of each ionic species, which allows the ionizing spectrum to be reconstructed. 

The nebular spectrum of any star forming galaxy is dominated by its most ionizing and hence massive stars. The blue ultraviolet spectra observed longwards of the Lyman break in typical high redshift galaxies are consistent with young stellar populations. However deep spectra of distant galaxies have long hinted that the role of massive stars might not be fully understood. Early work on the spectra of Lyman break galaxy samples at $z\sim3-4$ established that the absorption lines associated with stellar photospheres were offset in velocity from resonance line emission, including the Lyman-$\alpha$ emission line, with typical offset velocities of several hundred kilometers per second \citep{2003ApJ...588...65S}. The galaxy-scale outflows inferred from these velocity offsets prove ubiquitous in the distant Universe and require a substantial energy injection into the interstellar medium - either as a result of stellar winds from massive stars or a high CCSN rate associated with the starburst. 

A more direct, but harder to interpret, indication of massive star populations was first observed in a deep ultraviolet spectrum obtained by stacking spectroscopy of 811 Lyman break galaxies \citep{2003ApJ...588...65S}. This spectrum showed a number of broad, high ionization ultraviolet absorption features associated with massive O stars including the Si\,{\sc IV} and C\,{\sc IV} doublets, and - more surprisingly - a velocity broadened \heii\ recombination line at a rest-frame wavelength of 1640\,\AA\ in emission. 
\heii\ emission requires a source of photons with an energy of 54.4eV, i.e. a  $\sim53$\,kK photosphere, hotter than expected from main sequence stars. In the local Universe both broad \heii\ and narrow nebular emission was expected to arise primarily from Wolf-Rayet stars or their environs  \citep[e.g.][]{1998ApJ...497..618S}. But in most stellar populations, the number of WR stars relative to O-stars is small and 
it is  rare for \heii\ emission to be detected in the integrated light of galaxies with any significant strength. To see it in a stack implied a substantial fraction of the galaxy population at $z\sim3$ must host an abundant WR population that appeared inconsistent with stellar population model fits to the whole spectrum \citep{2003ApJ...588...65S}. Proposed explanations included the possibility that unenriched primordial gas might allow Population III star formation to continue as late as 2-3 Gyr after the Big Bang \citep{2006Natur.440..501J,2007MNRAS.382..945T} - a prospect which seemed unlikely given cosmological models and the rate at which the intergalactic medium was enriched, and also the broad line emission. Population III stars are not expected to drive fast, optically-thick  winds, due to their low envelope opacity, and so should generate a narrow nebular line emission  spectrum.



Since these early investigations there have been a large number of studies of both individual and stacked galaxies, exploiting the ever increasing sensitivity and wavelength coverage of modern spectrographs to explore both distant galaxies and local analogues which show similar ionization environments \citep[e.g.][]{2019ApJ...882..182C,2019ApJ...878L...3B,2018ApJ...859..164B,2020MNRAS.497.4293G}. 
In particular, the advent of highly-multiplexed near-infrared spectrographs - notably MOSFIRE on the 10m Keck telescope - have led to a rapid increase in the number of galaxies at redshifts between cosmic noon and cosmic dawn with rest-frame optical emission line spectra. These allow construction of diagnostic diagrams which have traditionally used to separate stellar irradiation from that due to accretion onto quasars \citep[particularly the \lbrack O\,III\rbrack/H$\beta$ vs \lbrack N\,II\rbrack/H$\alpha$ BPT diagram, ][]{1981PASP...93....5B}. Analysis of samples at $z\sim2-3$ indicate that the locus of star forming galaxies in these parameter spaces is offset towards a higher ionization potential than those in the distant Universe, indicating either a harder ionizing spectrum or unexpected elemental abundance variations in the key species \citep{2016ApJ...826..159S, 2017ApJ...836..164S,2018ApJ...868..117S}. This latter possibility can be somewhat controlled for by considering the ionization potential in a variety of different diagnostic spaces, strongly indicating that a harder ionizing radiation field is at work at $z>2$ than is typically seen locally \citep[e.g.][]{2020MNRAS.491.1427S,2021MNRAS.502.2600R}. Over the coming years, the James Webb Space Telescope will expand samples of galaxies with this information to higher redshift and permit more detailed resolved studies at around cosmic noon (see section \ref{sec:obs-outlook}).

\begin{marginnote}
\entry{Threshold Energies} \ 
\entry{Ly$\alpha$, H$\alpha$, H$\beta$ etc} \hfill 13.6\,eV
\entry{[O\,{\sc III}]} \hfill 35.1\,eV
\entry{[Ne\,III]} \hfill 40.96\,eV
\entry{[C\,III]} \hfill 47.9\,eV
\entry{He\,II} \hfill 54.4\,eV

\end{marginnote}

Returning to the rest-frame ultraviolet, in addition to both broad and narrow \heii\ emission, some of these sources show evidence for emission in other high ionization potential lines, including C\,{\sc III} 1909\,\AA, and C\,{\sc IV} 1550\,\AA\ seen in emission rather than absorption \citep[][]{2014MNRAS.445.3200S,2018A&A...616L..14S}. Similar evidence is now being seen in the far-infrared, where anomalously high [O\,{\sc III}]\,88\,$\mu$m / [C\,{\sc II}]\,158\,$\mu$m line emission ratios have been observed in the distant Universe \citep{2020ApJ...896...93H}. 
These hard ionizing spectra are by no means ubiquitous, but are sufficiently common that they must be a relatively normal product of stellar populations \citep[e.g.][]{2017A&A...608A...4M}. 
Importantly, there are no clear indications of AGN activity. These observations have built a picture in which there is substantial variation from galaxy to galaxy, but in which young stellar populations at low metallicity are typically characterised by a very hard, blue ionizing spectrum. 

Calculating quantitative constraints on the ionizing spectrum from these galaxies is not straightforward. The strength and equivalent width of most nebular emission lines is influenced not only by the ionizing spectrum but also by the density, geometry and composition of the nebular gas, as well as the escape fraction of photons as a function of energy.  As a result, lines such as  [O\,{\sc III}], [Ne\,III] and [C\,III] in emission (probing photons at 35.1\,eV, 40.96\,eV and 47.9\,eV respectively) are challenging to interpret without detailed fitting. This lies outside the scope of stellar population synthesis models, and requires post processing of the stellar emission by a radiative transfer codes \citep[e.g.][]{2017RMxAA..53..385F} 
able to calculate multiple ionization zones, incorporating comprehensive atomic and molecular data, and key grain physics. In particular, many heavy element forbidden lines are highly sensitive to density and electron temperature, as well as both overall metallicity and the relative abundance of different elements.

Exceptions are the recombination lines of hydrogen and helium, which are relatively insensitive to nebular conditions and well understood, and so can be translated directly to an ionizing photon production rate.
Such conversions assume that the escape fraction, luminosity distance and extinction in a galaxy is well understood, and thus a more robust observational quantity 
is a line ratio, particularly of the optical \heii\,4686\,\AA\ and H$\beta$\,4861\,\AA\ lines, yielding a ratio of Q(\heii)/Q(H0) = Q(55.4\,eV)/Q(13.6\,eV) photon fluxes. 

\begin{figure}
    \centering
    \includegraphics[width=0.91\textwidth]{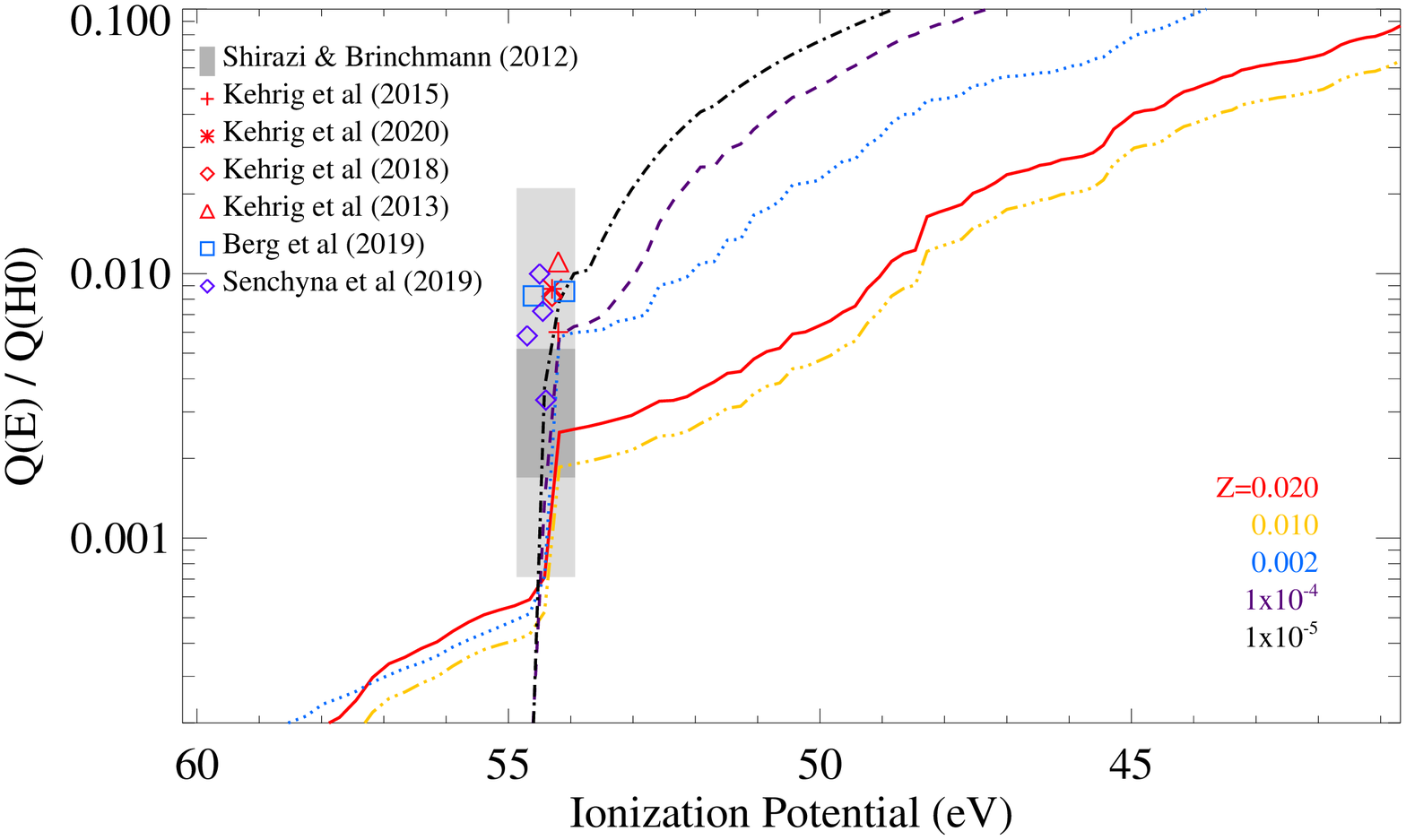} 
    \caption{The photon flux below a given energy threshold Q(E) as a function of ionization potential energy, normalised by the photon flux at 13.6eV (912\,\AA). Lines show five zero-age model stellar populations at different metallicities. The grey shaded region indicates Q(\heii)/Q(HI) ratios measured for Wolf-Rayet galaxies in the local Universe by  \citet{2012MNRAS.421.1043S}  - we indicate both the full range of measurements (light grey) and the mean and standard deviation (dark grey). Values derived from individual extreme emission line galaxies exhibiting \heii\ emission are shown as symbols \citep{2013MNRAS.432.2731K,2015ApJ...801L..28K,2018MNRAS.480.1081K,2020MNRAS.498.1638K,2019ApJ...878L...3B,2019MNRAS.488.3492S}. Uncertainties are typically of order $\sim$0.1-0.2\,dex. All measurements lie at 54.4\,eV and small random offsets are applied for clarity. Where ionizing photon rates are not calculated by the original authors, we adopt the calibrations given in the Supplementary Information.}
    \label{fig:qhs_data}
\end{figure}

In Figure \ref{fig:qhs_data} we demonstrate how the ionizing spectrum of a zero age starburst varies with metallicity, given our current understanding of the massive star population. At this age, the difference between single star and binary models is negligible. At later ages, the binary population boosts the number of helium ionizing photons, particularly at low metallicity \citep[see ][]{2019A&A...621A.105S}. The models demonstrate the rapid fall off in photon production with increasing energy, together with an edge at 54\,eV caused by helium ionization in the stellar photospheres. We also show observational constraints on the ratio of photons at this energy to those at 13.6\,eV, obtained from optical line ratios, for extreme emission line galaxies and Wolf-Rayet galaxies in the local Universe. 


Binary stellar evolution models have proved better able than single star models to reproduce \heii\ emission in the majority of the galaxy population both locally and at high redshift \citep{2009MNRAS.400.1019E,2012MNRAS.419..479E,2016ApJ...826..159S}. However evidence such as Figure \ref{fig:qhs_data} suggests that current model stellar populations do not generate sufficient hard photons 
explain the spectra of the most ionizing stellar populations and resolving this will not be straightforward. 
\citet{2019A&A...621A.105S} demonstrated that reasonable variation in the initial mass function was unable to resolve this difficulty, although there are still significant uncertainties 
in the true shape of the ionizing spectrum for hot stars which have been stripped of their envelopes, either by winds or binary interactions (see section \ref{sec:unc-pops}). 

\begin{marginnote}
No current SPS models are able to fully account for the hardest far-UV spectra seen in low metallicity galaxies. This hints at additional unknown or poorly understood sources of hard ionizing photons.
\end{marginnote}

A possible contributor is emission from accretion onto compact objects (either in the X-ray from massive binaries or the extreme-ultraviolet from lower mass systems). As figure \ref{fig:SFRIs} indicates, binary populations efficiently generate X-ray binaries as soon as the first compact objects form from massive stars. Galaxies at low metallicity show an elevated $L_X$/SFR \citep[see e.g.][]{2020MNRAS.495..771F,2021MNRAS.505.4798S,2020NatAs...4..159B}, and local low metallicity extreme starbursts show an excess of ultraluminous X-ray sources \citep{2016ApJ...818..140B} which may result from these stars retaining more of their mass until the end of their lifetime due to weaker stellar winds. However the role of X-ray binaries in the ionizing spectra of galaxies is still debated in the literature. \citet{2020MNRAS.496.3796S} have shown that \heii\ emitters do not seem to differ from other star forming galaxies at $z\sim3$ in their hard X-ray luminosity, while some models \citep[e.g.][]{2021arXiv210812438S} suggest that they may well be responsible for observed \heii\ emission lines, given certain assumptions. Importantly, work to date has focussed on the very hard spectra of high mass X-ray binaries, while lower mass binaries may also prove important. Constraints on the extreme-ultraviolet/supersoft-X-ray spectra of galaxies are weak, and the emitted spectrum may depend on a number of issues including the inner truncation radius (and hence maximum temperature) of accretion disks. This in turn is strongly sensitive to the magnetic fields of neutron star or white dwarf accretors.  Further work is needed to fully explore the consequences of binary and low metallicity massive star evolution in this regime.


\begin{summary}[SUMMARY POINTS]
\begin{enumerate}
\item The distant Universe provides a testbed for models of the impact of massive stars.
\item Identification of high ionizing photon production efficiencies in young galaxies, particularly at low metallicity, suggests that simple, single-star stellar population models underestimate the ionizing radiation field.
\item Modelling of reionization suggests that binary stellar populations boost the escape fraction of ionizing radiation by opening channels through the interstellar medium.
\item Spectroscopy of distant galaxies suggests that they have hard ionizing radiation fields, hinting at an overabundance of massive stars or at a mechanism to prolong the lives of those stars, such as binary interactions or potentially rotation.
\end{enumerate}
\end{summary}

\section{UNCERTAINTIES AND OUTLOOK}

\subsection{Key Uncertainties in Massive Star Evolution}

We can be optimistic that we now have a very good understanding of the general evolution of massive stars in early galaxies. Model stellar populations that take account of the full range of binary interactions,  using recent stellar atmosphere models, reproduce observations of early galaxies well. However, there remain areas disagreement between these model stellar populations and observations that require further study.

The impact of rotation, and in particular the differential rotation rates of stellar interiors and their atmospheres, and the impact of subsequent mixing, remains an active and fast-moving area of research \citep[see e.g.][]{2019ARA&A..57...35A}, even within isolated single massive stars. Nor is rotation the only source of uncertainty regarding mixing. Additional sources of stratified mixing may also be required by new insights into stellar interiors from asteroseismology \citep{2021NatAs...5..715P}. The relative importance of further rotation that results from binary interactions later in stellar evolution also remains unclear. The formation of QCHE stars from very rapid rotation seems to be required (from BPASS models) but local analogues have yet to be explicitly identified. 

The signature of the contribution of the energetic stellar deaths such as PISNe, long-GRBs and SLSNe to early galaxies has yet to be clearly identified. While rare events, these could have an important impact in the early evolution of early galaxies, for example by clearing out their environs so that the ionizing photons from less massive stars can escape into the IGM. The possible impact of the absence of supernovae from massive core-collapse also needs to be considered: this implies that we must rely on the rare and unusual extreme supernovae even more to clear out the star-forming environment. 

The evolution and impact of very massive stars also requires further understanding. These stars contribute a significant fraction of the ionizing photons from a young stellar population, and may also interact with their own winds to produce shocks that may explain some of the stronger emission lines that have been observed in early galaxies. Studying such stars, however, is difficult given their rarity in the local Universe.

\subsection{Key Uncertainties in Synthetic Populations}\label{sec:unc-pops}

Work to quantify the uncertainties in synthetic populations is an ongoing and iterative process, mostly carried out by comparison with data \citep[e.g.][]{2016MNRAS.457.4296W,2019A&A...621A.105S,2019ApJ...878....2D,2020MNRAS.495.4605S}. 
Some sources of uncertainty, such as a galaxy's star formation history, introduce an inevitable degeneracy with the stellar parameters and must by resolved by fitting. Others are intrinsic to the population models. There are, perhaps, two important areas in which significant room remains for improvement.

The first of these is our understanding of the binary fraction of stellar populations. At near-Solar metallicities in the local Universe this is relatively well constrained \citep{2017ApJS..230...15M}, and the remaining uncertainty has a small and quantifiable impact on outputs such as the ionizing photon production \citep{2020MNRAS.495.4605S}. However there are few constraints on the dependence of binary fraction with metallicity or other star formation properties and most binary population indicators are indirect or are qualatitive. If molecular cloud fragmentation is affected by composition, as seems likely, then binary fractions will be metallicity dependent. Current indications are that low metallicities may promote binary formation in low mass, Solar-type stars \citep{2019ApJ...875...61M}, but there are hints that the binary fraction of high mass stars may actually be lower \citep{2021arXiv210708304M}. However data remains very sparse and is likely to remain so until the advent of the Extremely Large Telescopes allows resolved spectroscopy of more distant stellar populations. There are also some indications of a dependence of binary fraction on enhancement of $\alpha$-process elements \citep{2020MNRAS.499.1607M}, although it is unclear whether this is independent of overall metallicity trends. This is an area which may be further explored as improved spectroscopy of massive binary signatures in distant galaxies becomes available.

A second major area in need of work is evaluating the uncertainties in the stellar atmosphere models used to generate synthetic population spectra. While the optical spectra of main sequence and giant stars are well understood, models tend to diverge shortwards of the $\sim$3500\AA\ atmospheric cut-off, where observational data is sparse. The ongoing ULYSSES project \citep{2020RNAAS...4..205R} aims to improve on this by generating a library of well-characterised stars with space-based ultraviolet spectroscopy. This can be used to tune models, but will still leave considerable room for uncertainty in the ionizing spectra of stars, emitted below the Lyman limit.

However far less work has been done on the spectra of binary products. In the case of many early mergers or after minor mass-transfer episodes, these may well be indistinguishable from those of normal stars. However stars which evolve quasi-homogeneously, or which have been stripped through binary interactions, occupy regions of temperature-gravity-luminosity-composition parameter space not populated by single-star models. 

For massive stars driving optically-thick winds (classical Wolf-Rayet stars) the leading theoretical models are generated by the PoWR code \citep[e.g.][and references therein]{2015A&A...579A..75T}. However at low metallicities and lower masses, envelope-stripping will likely lead to more compact helium stars, whose atmospheres have only recently been modelled in detail \citep{2017A&A...608A..11G,2018A&A...615A..78G}. 
These new atmospheres are not currently implemented in the BPASS models, and are currently only included as an option in the Starburst99 model grid \citep[which includes limited binary evolution pathways, ][]{2019A&A...629A.134G}. Importantly, such models cannot be directly calibrated in the local Universe, since these high energy photons are highly scattered by the interstellar medium, and thus there is scope for uncertainty in the microphysics of the atmosphere calculations. However any stellar atmosphere model is likely to encounter a significant problem: 
any star with a helium atmosphere will show the sharp absorption edge seen in the figure, and very few photons escaping at energies above the ionization threshold. Avoiding such absorption would require the exposure of a bare CO-core, which would have a vanishingly small lifetime, and is also challenging with binary stripping and likely impossible with low metallicity stellar winds. 

Work is also ongoing to assess the impact of composition differences (primarily $\alpha$-element enhancement) on stellar population models. A number of studies have now shown evidence for enhanced $\alpha$-element abundance in the young stellar populations of distant galaxies \citep[e.g.][]{2016ApJ...826..159S,2019MNRAS.487.2038C,2020MNRAS.491.1427S} - unsurprising since the old stellar populations seen in giant galaxies at the current epoch were forming the majority of their stars at $z>3$. While some progress is being made towards including $\alpha$-enhancement in stellar evolution, stellar population and spectral synthesis models \citep[e.g.][Byrne et al. in prep]{2018MNRAS.476..496F,2018ApJ...854..139C,2014MNRAS.440.1027C}, no current stellar evolution library currently considers both binary interactions or rotation and elemental composition. A lack of large, public and high resolution spectral libraries extending into the ultraviolet for stellar atmospheres at non-Solar compositions also presents a challenge for the SPS, and hence the galaxy evolution, community.  As an intermediate measure, some authors opt to use solar-scaled abundance models for the stellar population (whose emission is largely determined by iron opacity) while applying $\alpha$-enhancement to the nebular gas used in radiative transfer models \citep[since nebular emission includes strong features from $\alpha$-elements, e.g.][]{2017ApJ...836..164S}. While this is a useful compromise, it is important to note that the stars and their surrounding nebular gas are expected to have the same intrinsic composition.

\subsection{Theoretical Outlook}\label{sec:theory-outlook}

Progress is being made on the refinement of mass-loss rates for hot stars, both on the main sequence \citep[e.g.][]{2021A&A...648A..36B,2021MNRAS.504.2051V,2021arXiv210808340H} and for Wolf-Rayet stars (see Vink, this volume). However these remain uncertain. 
Further clarification of mass loss rates is important both to understand the evolution of these stars and how much of the mass loss must come from binary interactions. An area that is not being studied enough is the mass-loss rates of red supergiants. These rates are uncertain, with  \citet{2020MNRAS.492.5994B} finding they may need significant revision compared to the widely used rates of \citet{1988A&AS...72..259D}. While these stars do not have much impact on their environments themselves, they are the possible progenitors of WR stars.

The mass of the remnants that form in core-collapse is also a significant area of  research. Many recent studies suggest that black hole-forming events don't emit supernovae, and it remains uncertain exactly how massive black holes are when they are formed in core-collapse.  Remnant mass distributions are important both for GW transients and XRB population studies.

Finally the uncertain aspects of rotation and binary interactions are the subject of many ongoing studies and likely will continue to be so for the foreseeable future. While these are clearly important, exactly how a star's evolution changes while and after it interacts is still unclear. More detailed models of binary evolution, exploring more pathways and initial parameters, and probing their underlying physics, are required, in contrast to the approximate rapid models used by some current population synthesis codes. Only when we are able to have greater confidence that the binary pathways that are modelled reflect real binary stars will we be able to apply them to understand galaxies more accurately. We note that any model that explains the electromagnetic observations of these early galaxies must also explain the growing population of gravitational wave transients, some of whose progenitors were once stars in early galaxies.

\subsection{Observational Outlook}\label{sec:obs-outlook}

The observational outlook for this area is 
to a large extent driven by new facilities, as well as new techniques. 

The Atacama Large Millimeter Array (ALMA) has started to generate a growing catalogue of galaxies with detections of far-infrared emission lines and continuum  \citep[e.g.][]{2021MNRAS.505.5543V,2019PASJ...71...71H}. The continuum can provide stronger constraints on the dust temperature and properties in distant galaxies than has hitherto been possible, while the far-infrared line emission will complement that in the ultraviolet and near-infrared, allowing stronger constraints on both the ionizing spectrum and abundance ratios in distant sources. Importantly, ALMA can achieve spatial resolutions comparable to those of optical and near-infrared facilities, and so has the potential to reveal internal structure and variations between components of individual galaxies, particularly when coupled with strong lensing. Such work has already begun \citep[e.g.][]{2021ApJ...911...99F}, but its potential is yet to be fully embraced by modellers and will likely have a growing impact as sample sizes grow at intermediate to high redshifts, observations become more sensitive, and models are challenged to explain them.


Moving from an extant telescope to an imminent one, the field eagerly awaits the advent of the James Webb Space Telescope (JWST). JWST will permit optical imaging of galaxies extending into the tail end of the epoch of reionization. This will provide important constraints on the older stellar populations and star formation histories of galaxies within the first two billion years of the onset of galaxy formation. It also has the potential to reveal ultraviolet detections of more distant galaxies, potentially probing star formation at the peak of the reionization epoch. Unfortunately, it is likely that many of the faintest or most distant of these may prove too dim for spectroscopic follow-up, 
at least until the era of the Extremely Large Telescopes and future space telescope missions.

Perhaps more exciting is the spectroscopic capability of JWST. Both the rest-frame ultraviolet and the rest-frame optical will be accessible for galaxies in the distant Universe, at unprecedented spectral resolution and signal-to-noise. This will permit line ratio and ionization potential analyses that have enhanced our understanding of stellar populations at cosmic noon to be extended towards cosmic dawn, and confirm or refute many of the interpretations that have been proposed for early galaxy emission features. Crucially these will include signatures of chemical enrichment, and of the strong ionizing fields of massive stars in an age, density, metallicity and star formation density regime inaccessible in the local Universe. We fully anticipate that these will challenge models. Perhaps most interesting will be direct measurements of the ionizing photon production rates, escape fractions and ionizing spectra from sources in the process of reionizing the Universe, rather than their analogues and descendants. The published program for JWST already includes both guaranteed time observation and general observer surveys which aim to secure large samples of such galaxies.

The prospect for observing the very first stars is perhaps a little more remote. While Population III signatures may be seen in spectroscopy of galaxies during cosmic dawn, until the origin of hard ionizing radiation fields in Population II is understood, these signatures may be difficult to separate from those of massive stars in the underlying population. They are also likely to be relatively short-lived events, occurring in pockets of primordial gas in galaxies already somewhat enriched by star formation. There is perhaps some hope of observing transients attributable to Population III events, either as well-localised gravitational wave transients or very distant gamma-ray bursts. Unfortunately the latter possibility has suffered a recent setback with the deselection of the proposed Theseus space telescope  \citep{2021ExA...tmp...97T}, and it is to be hoped that future opportunities will arise. Given the interest in the community regarding massive stars in distant galaxies, and the renewal of interest in the infrared which will  follow from the launch of JWST and the subsequent Nancy Grace Roman Space Telescope, this remains a good possibility.

%

\section{CONCLUSIONS}
\begin{summary}[SUMMARY POINTS]
\begin{enumerate}
\item Theoretical understanding of the evolution of massive stars, and their impact on their surroundings, has improved significantly in recent years, although regions of uncertainty have become clearer.
\item Simple stellar populations that account for only single stars are currently unable to explain observations of early galaxies. The stellar populations must include the full range of binary evolutionary pathways to begin to explain these early galaxies. 
\item Observations of early galaxies that will aid understanding of massive star processes include the spectrum shape of the Lyman continuum, detection of supernovae at high redshift and large samples of observed young and low metallicity galaxies.
\end{enumerate}
\end{summary}
\begin{issues}[FUTURE ISSUES]
\begin{enumerate}
\item An improved understanding is needed of the initial strength, evolution and impact of stellar rotation within both single and binary stars is required. No current SPS code currently accounts in detail for both rotation and stellar multiplicity.
\item Some initial parameters of stellar populations, such as the distribution of very high mass stars and the dependence of binary parameters on metallicity, remain unclear.
\item The role of X-ray binaries in modifying galaxy light must be further explored.
\item Exactly how the most massive (M$>100$\,M$_\odot$) stars evolve and die may be key to understanding the youngest stellar populations. 
\item The role of elemental composition (particularly [$\alpha$/Fe] ratio) in the properties of massive stars remains to be fully considered in most population synthesis models.
\item The origin of the hardest photons seen in young, low metallicity star forming galaxies is unclear and may require revised far-ultraviolet stellar atmosphere models or new sources.
\end{enumerate}
\end{issues}

\section*{DISCLOSURE STATEMENT}
The authors are not aware of any affiliations, memberships, funding, or financial holdings that might be perceived as affecting the objectivity of this review. 

\section*{ACKNOWLEDGMENTS}
The authors thank the editor and referees for comments that improved the review. The authors thank past and present BPASS team members for fruitful collaboration over many years. JJE acknowledges support by the University of Auckland and funding from the Royal Society Te Apar\={a}ngi of New Zealand Marsden Grant Scheme. ERS acknowledges support by the University of Warwick and funding from the UK Science and Technology Facilities Council (STFC) through Consolidated Grant ST/T000406/1.

%


\bibliographystyle{ar-style2}
\bibliography{bib}

\newpage
\renewcommand\theequation{A\arabic{equation}}
\setcounter{equation}{0}
\renewcommand\thetable{A\arabic{table}}
\setcounter{table}{0}
\section*{SUPPLEMENTARY INFORMATION}

In this supplementary information we give calibrations for ionizing photon production rate as a function of observed recombination line luminosities, as used in Sections \ref{sec:xiion} and \ref{sec:spec}. This uses recombination constants tabulated in \citet[][hereafter OF06]{2006agna.book.....O} and closely follows that synthesis.

For Case B recombination (optically thick, appropriate for most extragalactic nebulae), the ionizing photon rate in a species X relates to its recombination lines as:

\begin{equation}
Q(\mathrm{X}) = \mathrm{V}\,n_e\,n(\mathrm{X}^+)\,\alpha_B(\mathrm{X}) = \mathrm{V}\,\frac{4\,\pi\,j_{nn'}}{h\,\nu_{nn'}}\frac{\alpha_B(\mathrm{X})}{\alpha^\mathrm{eff}_{nn'}} = \frac{L_{nn'}}{h\,\nu_{nn'}}\frac{\alpha_B(\mathrm{X})}{\alpha^\mathrm{eff}_{nn'}}
\end{equation}

Here V is the volume of the emitting region and $j_{nn'}$ is the emissivity of the line corresponding to the $n' \rightarrow n$ transition, which is emitted at a frequency $\nu_{nn'}$. $L_{nn'}$ is the line luminosity. $\alpha_B(\mathrm{X})$ is the temperature-dependant case B recombination constant appropriate for the species, and $\alpha^\mathrm{eff}_{nn'}$ is the temperature- and electron density-dependant effective recombination constant for the line defined by $\alpha^\mathrm{eff}_{nn'} = 4\,\pi\,j_{nn'} / n_e\,n(\mathrm{X^+})\,h\,\nu_{nn'}$\\
\\
We make use of tables 2.4 and 2.7 of OF06 to determine:

\smallskip
\begin{tabular}{lcc}
    Species &  $\alpha_B$(T=10,000K) & $\alpha_B$(T=20,000K)  \\
    \hline
     $\alpha_B(\mathrm{H}^0$)  & $2.59 \times 10^{-13}$ &  $1.43 \times 10^{-13}$ cm$3$ s$^{-1}$\\
     $\alpha_B(\mathrm{He}^0$) = 2\,$\alpha_B(\mathrm{H}^0$,T/4) & $2.72 \times 10^{-13}$ &  $1.56 \times 10^{-13}$ cm$3$ s$^{-1}$\\
     $\alpha_B(\mathrm{He}^+$) & $1.54 \times 10^{-12}$ &  $9.08 \times 10^{-13}$ cm$3$ s$^{-1}$\\
\end{tabular}
\\
\\

OF06 tabulates effective line recombination constant values for reference lines in each species, and provides a ratio of emissivities for additional lines, which allows their effective recombination constants to also be determined:

\begin{equation}
    \alpha^\mathrm{eff}_{nn'} = \frac{j_{nn'}}{\nu_{nn'}}\frac{\nu_\mathrm{ref}}{j_\mathrm{ref}}\,\alpha^\mathrm{eff}_\mathrm{ref} = \frac{\lambda_{nn'}}{\lambda_\mathrm{ref}}\frac{j_{nn'}}{j_\mathrm{ref}}\alpha^\mathrm{eff}_\mathrm{ref}
\end{equation}\\
\\

We now use values provided in tables 4.4, 4.5 and 4.6 of OF06 to derive conversion constants of the form  $Q(\mathrm{X}) = A\,L_\mathrm{line}$ where Q is in photons s$^{-1}$, $L_\mathrm{line}$ is the line luminosity in ergs s$^{-1}$ and we give values at 10,000 and 20,000\,K for $n_e = 100$\,cm$^{-3}$. These conversion factors are tabulated in Table \ref{tab:alpha}.

\begin{table}[h]
\tabcolsep7.5pt
\caption{Line Luminosity to Ionizing flux conversions, for $n_e=100\,cm^{-3}$ and T=10,000\,K or T=20,000\,K, derived from \citet{2006agna.book.....O}.}
\label{tab:alpha}
\begin{center}
\begin{tabular}{@{}l|c|c|c|c@{}}
\hline
Atomic Species, line & $\lambda$ & Transition & A(10kK)$^{\rm b}$ & A(20kK)$^{\rm b}$\\
 & \AA &  & (phot s$^{-1}$ / ergs s$^{-1}$) &(phot s$^{-1}$ / ergs s$^{-1}$)\\
\hline
$H^0$, H$\alpha$&     6563 &   $ 3\rightarrow2$  &      $1.28\times10^{12}$ &      $1.43\times10^{12}$\\
$H^0$, H$\beta$ &     4861 &   $ 4\rightarrow2$  &      $3.68\times10^{12}$ &      $3.93\times10^{12}$\\
$He^+$, \heii\   &     1640 &   $ 3\rightarrow2$  &      $1.57\times10^{11}$ &  $1.69\times10^{11}$\\
$He^+$, \heii\   &    4686  &   $ 4\rightarrow3$  &      $1.02\times10^{12}$ &  $1.24\times10^{12}$\\
$He^0$, He I$^{\rm a}$     &    3889 &    3\, $^1S\rightarrow2\,^3S$  &      $1.91\times10^{12}$ &      $1.86\times10^{12}$\\
$He^0$, He I     &    4471 &    4\,$^3D\rightarrow2\,^3P^0$ &       $4.40\times10^{12}$&       $5.14\times10^{12}$\\
$He^0$, He I     &    5876 &    3\,$^3D\rightarrow2\,^3P^0$ &       $1.66\times10^{12}$&       $1.96\times10^{12}$\\
$He^0$, He I     &    6678 &    4\,$^1D\rightarrow2\,^1P^0$ &       $5.82\times10^{12}$&       $7.13\times10^{12}$\\
$He^0$, He I     &    7065 &    3\,$^3S\rightarrow2\,^3P^0$ &       $9.00\times10^{12}$&       $7.18\times10^{12}$\\
$He^0$, He I     &    10830 &   2\,$^3S\rightarrow3\,^3P^0$ &       $8.14\times10^{11}$&       $6.60\times10^{11}$\\
\hline
\end{tabular}
\end{center}
\begin{tabnote}
$^{\rm a}$Constants for He\,I do not include radiative transfer effects \\
$^{\rm b}$Conversion factors in the form $Q(\mathrm{X}) = A\,L_\mathrm{line}$ where Q(H$^0$), Q(He$^+$) and Q(He$^0$) are in phot\,s$^{-1}$, $L_\mathrm{line}$ is the appropriate line luminosity in ergs\,s$^{-1}$.
\end{tabnote}
\end{table}

\end{document}